\documentclass[review, 3p, 8pt, times, authoryear,fleqn]{elsarticle}
\usepackage{amsmath,amssymb,bm}%cleverref must be loaded after this
\usepackage[colorlinks=true,citecolor=blue]{hyperref}% for blue color links
\usepackage[nameinlink,capitalise]{cleveref} %clever ref to call figures
\usepackage{ulem} % For striking out text
\journal{Additive Manufacturing}  
\makeatletter
\def\ps@pprintTitle{%
  \let\@oddhead\@empty
  \let\@evenhead\@empty
  \let\@oddfoot\@empty
  \let\@evenfoot\@oddfoot
}
\makeatother % To remove preprint submitted to a journal

\begin{document}
\begin{frontmatter}
% \title{Estimating the bounds on anisotropic elastic moduli in two-dimensional structured materials}
\title{Planar structured materials with extreme elastic anisotropy}
\author[1]{Jagannadh Boddapati}
\author[1]{Chiara Daraio}
\ead{daraio@caltech.edu}   
\cortext[cor1]{Corresponding author}
% \fntext[ec]{These authors contributed equally.}
 %% Authors' Addresses
\address[1]{Division of Engineering and Applied Science, California Institute of Technology, Pasadena, CA 91125, USA}
%%%%%%%% ORCIDS %%%%%%%%%%%%%%%%%%%
%ORCID[0000-0001-8706-5963] Jag pboddapa@caltech.edu
%ORCID[0000-0001-5296-4440] Chiara  daraio@caltech.edu
%%%%%%%%%%%%%%%%%%%%%%%%%%%%%%%%%%%%%%%%%%
%%%%%%%%%% Abstract %%%%%%%%%%%%%%%%%%%%%%
%%%%%%%%%%%%%%%%%%%%%%%%%%%%%%%%%%%%%%%%%%
\begin{abstract}
Designing anisotropic structured materials by reducing symmetry results in unique behaviors, such as shearing under uniaxial compression or tension. 
This direction-dependent coupled mechanical phenomenon is crucial for applications such as energy redirection. 
While rank-deficient materials such as hierarchical laminates have been shown to exhibit extreme elastic anisotropy, there is limited knowledge about the fully anisotropic elasticity tensors achievable with single-scale fabrication techniques. 
No established upper and lower bounds on anisotropic moduli achieving extreme elastic anisotropy exist, similar to Hashin-Shtrikman bounds in isotropic composites. 
In this paper, we estimate the range of anisotropic stiffness tensors achieved by single-scale two-dimensional structured materials.
To achieve this, we first develop a database of periodic anisotropic single-scale unit cell geometries using linear combinations of periodic cosine functions.
The database covers a wide range of anisotropic elasticity tensors, which are then compared with the elasticity tensors of hierarchical laminates.
Through this comparison, we identify the regions in the property space where hierarchical design is necessary to achieve extremal properties.
We demonstrate a method to construct various 2D functionally graded structures using this cosine function representation for the unit cells. 
These graded structures seamlessly interpolate between unit cells with distinct patterns, allowing for independent control of several functional gradients, such as porosity, anisotropic moduli, and symmetry.
The graded structures exhibit unique mechanical behaviors when designed with unit cells positioned at extreme parts of the property space. 
Specific graded designs are numerically studied to observe behaviors such as selective strain energy localization, compressive strains under tension, and localized rotations. 
The designs exhibiting localized rotations under tensile loading are further experimentally validated through tensile tests on additively manufactured specimens.
%%%%%%%%%%%%%%%%%%%%%%%%%%%%%%%%%%%%%%%%%

\end{abstract}
%%%%%%%%%%%%%%%%%%%%%%%%%%%%%%%%%%%%%%%%%
%%%%%%%%% Graphical Abstract %%%%%%%%%%%%
% %%%%%%%%%%%%%%%%%%%%%%%%%%%%%%%%%%%%%%%%%
% \begin{graphicalabstract}
% % \includesvg[width=\textwidth]{img/graphical_abstract_v2.svg}
% \includegraphics[width=\textwidth]{img/graphical_abstract_v6.jpg}
% \end{graphicalabstract}
%%%%%%%%%%%%%%%%%%%%%%%%%%%%%%%%%%%%%%%%
%%%%%%%%%% Highlights %%%%%%%%%%%%%%%%%%
%%%%%%%%%%%%%%%%%%%%%%%%%%%%%%%%%%%%%%%%
% \begin{highlights}
% \item Estimating the bounds on the anisotropic elastic moduli of single-scale structured materials. 
% \item Identifying the regions in the elastic property space that require hierarchical designs.
% \item Design of functionally graded anisotropic structures with seamless transition between unit cells with distinct patterns. 
% \item Exploiting functional gradients to demonstrate atypical mechanical behavior such as localized rotations. 
% \end{highlights}
%%%%%%%%%%%%%%%%%%%%%%%%%%%%%%%%%%%%%%%%
%%%%%%%% Keywords %%%%%%%%%%%%%%%%%%%%%%
%%%%%%%%%%%%%%%%%%%%%%%%%%%%%%%%%%%%%%%%
\end{frontmatter}

%%%%%%%%%%%%%%%%%%%%%%%%%%%%%%%%%%%%%%%%%
%%%%%%%%% Main article %%%%%%%%%%%%%%%%%%
%%%%%%%%%%%%%%%%%%%%%%%%%%%%%%%%%%%%%%%%%

%%%%%%%%%%%%%%%%%%%%%%%%%%%%%%%%%%%%%%%%%
%%%%%%%%% Introduction %%%%%%%%%%%%%%%%%%
%%%%%%%%%%%%%%%%%%%%%%%%%%%%%%%%%%%%%%%%%
% \newpage
\section{Introduction}

%%%%%%%%%%%%%%%%%%%%%%%%%%%%%%%%%%%%%%%%%
%%%%%%%%% Metamaterials tensors  %%%%%%%%
%%%%%%%%%%%%%%%%%%%%%%%%%%%%%%%%%%%%%%%%%
Structured materials are engineered materials that derive special functionality from their micro- and meso- architecture. 
By fine-tuning the micro- and meso-architecture of both periodic and aperiodic tilings, a diverse range of effective mechanical properties can be achieved beyond what is possible with the corresponding material used for fabrication \citep{surjadi_mechanical_2019}.
As a result, the structured materials help achieve mechanical properties such as negative Poisson ratio \citep{saxena_three_2016} that their corresponding base materials could not achieve.
Among several mechanical properties of the metamaterials, elasticity tensors provide crucial information related to the energy density stored in the material, and indicate the directions in which the structure is most resistant or compliant to the applied loads. 
Using structured materials to tune local mechanical properties in the form of elasticity tensors \citep{panetta_elastic_2015, schumacher_microstructures_2015}, it is possible to design mechanical cloaks \citep{buckmann_elasto-mechanical_2014, wang_mechanical_2022}, artificial bone scaffolds \cite{gupta_artificial_2023}, cardiac stents \citep{wu_mechanical_2018}, wearable haptic interfaces \cite{oh_easy--wear_2023}, and elastic wave manipulating devices \citep{lee_elastic_2023}. 

%%%%%%%%%%%%%%%%%%%%%%%%%%%%%%%%%%%%%%%%%
%%%%%%%%% Anisotropy Applications %%%%%%%
%%%%%%%%%%%%%%%%%%%%%%%%%%%%%%%%%%%%%%%%%
Structured solids are not a new concept. Foams, for example, also derive their properties from their mostly hollow microstructure. 
However, due to the stochastic nature of the foams’ geometries, their elastic properties are isotropic and mostly dependent on the volume fraction of the material \citep{gibson_cellular_1997}. 
With advances in additive manufacturing, engineers design structured solids beyond foams and architect the material systematically to get desired mechanical behavior by decoupling the strong dependence on the volume fraction.
The designs range from simple truss-based lattice materials \citep{fleck_micro-architectured_2010} to intricate spinodoid metamaterials with arbitrary curvatures \citep{soyarslan_3d_2018}, frequently engineered to exhibit near-isotropic behavior characterized by two elasticity parameters.
However, when the elastic properties of structured materials become direction-dependent, more descriptors (elasticity tensor moduli) are required.
The mechanical behavior of a completely anisotropic material is described by 6 independent elasticity tensor moduli in two dimensions (2D) and 21 moduli in three dimensions (3D), as opposed to 2 when the properties are direction-independent. 
Such a high number of independent elastic moduli expands the design space of structured materials.
Thus anisotropy allows for coupled deformations, such as materials that can shear under uniaxial compression \citep{karathanasopoulos_mechanics_2020} and shear-shear coupling: where shear deformation in one direction induces shear stresses in a perpendicular direction. 
These coupled deformations have applications in shape-morphing \citep{agnelli_design_2022}, mode-conversion between longitudinal and shear elastic waves \citep{lee_polarization-independent_2024}, impact mitigation via energy redirection \citep{tomita_transition_2023} and sound attenuation \citep{liu_elastic_2016}. 
{Despite many studies on shear-normal coupled deformations, a comprehensive understanding of their fundamental limits remains inadequate.}

%%%%%%%%%%%%%%%%%%%%%%%%%%%%%%%%%%%%%%%%%%%%%%%%%%%%%%%%%
%%%%%%%%% Design Approaches and Limitations %%%%%%%%%%%%%
%%%%%%%%%%%%%%%%%%%%%%%%%%%%%%%%%%%%%%%%%%%%%%%%%%%%%%%%%
Designing metamaterials for anisotropy, despite their ability to attain unique properties, is a challenging task.
Besides the fact that fewer or no symmetries in the unit cell design lead to a higher degree of anisotropic properties, the factors that contribute to strong anisotropic behavior with coupled deformations remain largely unexplored.
Inverse techniques, such as topology optimization, are frequently used to obtain unit cells that possess desirable elasticity tensor \citep{sigmund_systematic_2009,diaz_designing_2003}. 
However, such inverse design techniques may not always be the best approach, as it is not known a priori whether the prescribed elasticity tensor is compatible with the provided geometric parametrization.
An alternative design approach involves predefined parametrization for the unit cell. 
This parametrization is then used to compute pre-computed databases, enabling the derivation of useful structure-property relationships 
\citep{panetta_elastic_2015,schumacher_microstructures_2015,ostanin_parametric_2018,lumpe_exploring_2021,luan_data-driven_2023,zhang_computational_2023,li_customizable_2024}.
These data-driven methods generate databases that function as quick reference tables, such as for identifying extremal structures at the boundary of the property space, and can serve as initial guesses for topology optimization \citep{chen_computational_2018}. 
They also provide robust datasets for machine learning-based design algorithms \citep{zheng_deep_2023,lee_data-driven_2024}. 
For example, using the latent space provided by variational autoencoders (VAEs), \cite{wang_deep_2020} demonstrated how to obtain complex topological and mechanical interpolations in various unit cells. 
Similarly, \cite{mao_designing_2020} used generative adversarial networks (GANs) to discover new stiffer unit cells beyond the training data.
By leveraging the gradients provided by artificial neural network models, it is now possible to perform inverse designs customized to specific anisotropic elasticity tensors \citep{kumar_inverse-designed_2020, bastek_inverting_2022, zheng_unifying_2023} and further to tailor nonlinear mechanical behavior \citep{yang_using_2019, maurizi_inverse_2022, bastek_inverse_2023}.
While these unique approaches can identify unit cells with diverse elasticity tensors beyond the isotropic class, %\citep{xu_design_2016}, 
the range of achievable properties is primarily constrained by the selected input design representation.
Additionally, most of these approaches are restricted to orthotropic elasticity, where coupled deformations are absent.
There is limited knowledge about the range of elasticity tensors, especially concerning the extent to which shear-normal deformations can be coupled in two-dimensional single-scale structured materials.

%%%%%%%%%%%%%%%%%%%%%%%%%%%%%%%%%%%%%%%%%
%%%%%%%%% Anisotropic Bounds %%%%%%%%%%%%
%%%%%%%%%%%%%%%%%%%%%%%%%%%%%%%%%%%%%%%%%
On one hand, achieving a complete characterization of the parameter space in terms of the moduli becomes exceedingly difficult in the absence of a unique parametrization of the input geometry. 
On the other hand, defining how close the obtained designs are to the theoretical bounds is also challenging, as little is known about the theoretical limits of anisotropic elastic moduli (no known closed-form expressions) \citep{milton_open_2021}.
This range of all possible effective tensors is mathematically known as \textit{G-closure}. 
In classical studies on isotropic composites, \cite{hashin_note_1961,hashin_variational_1962} present a variational approach to determine the upper and lower bounds on the effective bulk and shear moduli ($\kappa^*$ and $\mu^*$) by decomposing the elastic energy into hydrostatic and deviatoric parts. 
Hence, the bulk and shear moduli directly represent the energy that can be stored in the composites under hydrostatic and deviatoric loading respectively.
However, individual parameters in the fully anisotropic case do not hold such straightforward interpretations. 
Even in the isotropic case, there are areas in the property space of $\mu^*$ defined by theoretical bounds where composites are yet to be discovered \cite{milton_open_2021}. 
\cite{willis_bounds_1977,milton_variational_1988,cherkaev_coupled_1993,allaire_optimal_1993} extended the theory of Hashin-Shtrikman isotropic bounds to orthotropic elasticity tensors, by introducing  ``trace-bounds''. 
These trace bounds, categorized as 'bulk modulus type' and 'shear moduli type', are derived by bounding the trace of the inverse stiffness tensor (compliance tensor) projected onto specific tensor subspaces.
These calculations only establish bounds on a certain combination of elastic moduli.
Moreover, extending this method to fully anisotropic media with shear-normal coupling is non-trivial. 
This is because the energy cannot be decomposed into simple tensor components but involves a combination of them for any type of loading. 

To bridge this gap, \cite{milton_near_2018} expanded upon the theories proposed by \cite{willis_bounds_1977,allaire_optimal_1993} on isotropic composites to explore bounds on arbitrary stress-strain pairs in the anisotropic composites. 
The key finding from this work is that sequentially layered laminates are shown to achieve these bounds on stress-strain pairs.
By examining the sum of energy and complementary energies, they show that integrating a rank-deficient material, such as a pentamode me tamaterial within hierarchical laminates enables the attainment of extremality in the stress-strain space.
Pentamode materials were first introduced by \cite{milton_which_1995}, suggesting that using pentamode materials as fundamental building blocks allows for the achievement of arbitrary effective anisotropic properties.
Generally, elasticity tensors of extremal materials exhibit rank deficiency \citep{sigmund_new_2000,hu_engineering_2023}, a characteristic shared by pentamode materials and hierarchical laminates.
Additionally, hierarchical laminates also emerge as energy-minimizing optimal structures in various microstructure evolution problems \citep{freidin_new_2007,chenchiah_relaxation_2008,antimonov_phase_2016,freidin_two-phase_2019}, as they are shown to achieve constant stress or strain in one of the phases and leading to the optimization of the translation bounds \citep{francfort_sets_1994}.

%%%%%%%%%%%%%%%%%%%%%%%%%%%%%%%%%%%%%%%%%
%%%%%%%%% Problem at Hand %%%%%%%%%%%%%%%
%%%%%%%%%%%%%%%%%%%%%%%%%%%%%%%%%%%%%%%%%
While hierarchical structures and other rank-deficient materials could serve as a design guide in realizing extreme elastic anisotropy, physically realizing such structures demands advanced fabrication techniques that are still in development.
There is limited knowledge about the elasticity tensors achievable with single-scale fabrication techniques.
In this paper, we address this gap by sampling a diverse database of anisotropic unit cells created by combining periodic cosine functions of varied spatial frequencies, as discussed in \cref{sec: database generation}.
The database properties are then compared with the properties of hierarchical laminates in \cref{sec: rd} for the first time (to the best of our knowledge) which are considered as theoretical bounds. 
This comparison helps identify the regions in the property space where hierarchical designs necessary to achieve extreme elastic anisotropy.
In \cref{sec: grad}, we demonstrate a method to construct various functionally graded structures using this cosine function representation for the unit cells.
{We show that the graded structures seamlessly interpolate between unit cells with distinct patterns.}
We then analyze the mechanical behavior of two such graded structures to achieve energy localization and strain localization, utilizing unit cells with extreme anisotropy in the graded design.
Finally, we provide concluding remarks in \cref{sec: concl}. 

%%%%%%%%%%%%%%%%%%%%%%%%%%%%%%%%%%%%%%%%%
%%%%%%%%% Database Generation %%%%%%%%%%%
%%%%%%%%%%%%%%%%%%%%%%%%%%%%%%%%%%%%%%%%%
% \newpage
 \section{Generating a diverse unit cell database}
 \label{sec: database generation}
\begin{figure}[!htb]
	\begin{center}
		\centering 
		\includegraphics[width = 0.95\textwidth]{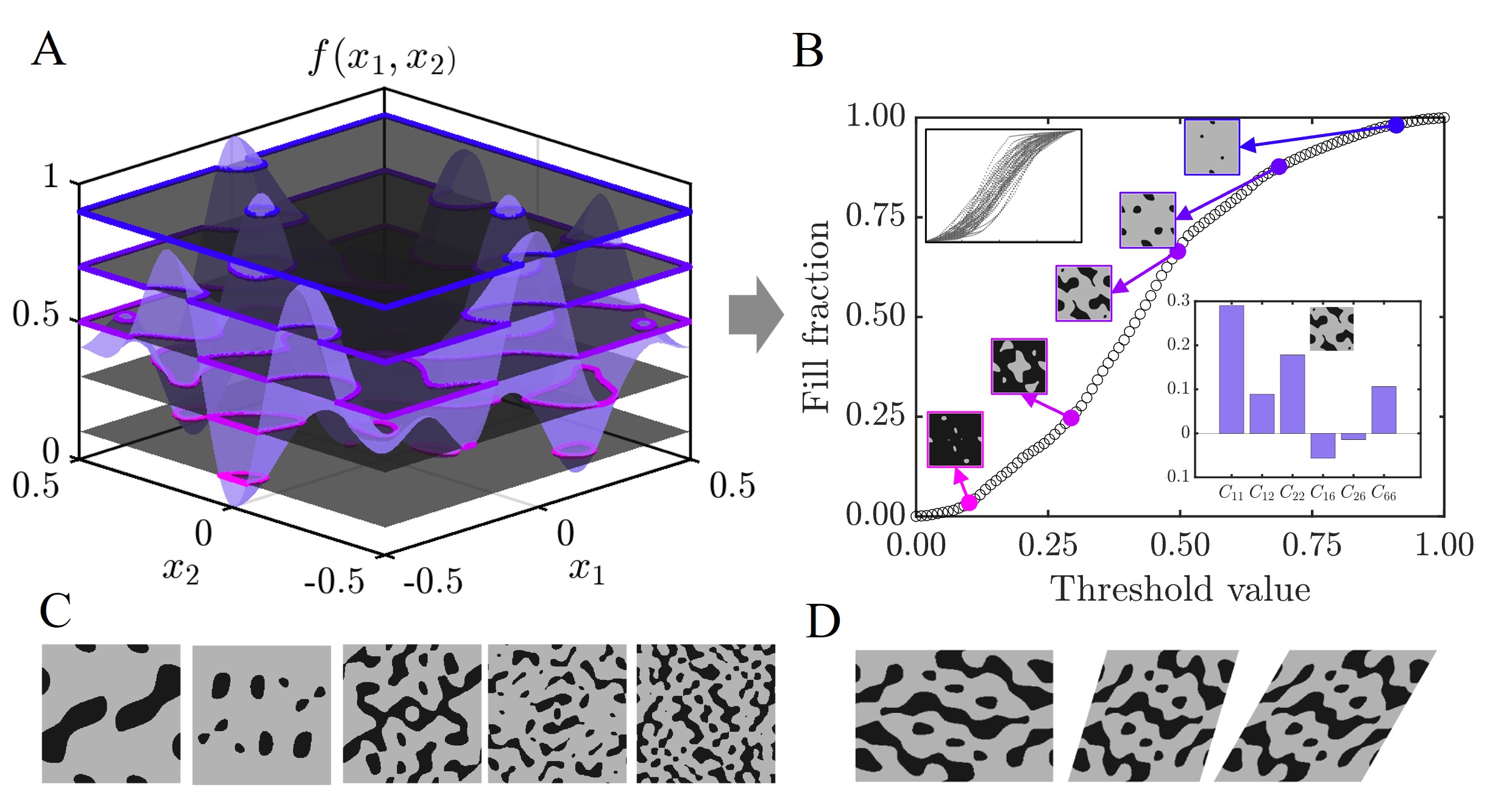}
		\caption{(A) Various anisotropic unit cells can be sampled by thresholding periodic functions composed of several cosine spatial modes, (B) {Variation of unit cell patterns with the fill fraction of the stiff phase as the threshold value is changed from 0 to 1 for a particular periodic function. The elastic properties of a unit cell normalized with Young's modulus of the stiff phase (bottom-inset), and the variation of the threshold-fill fraction curve for different realizations of the periodic function (top-inset).} (C) As the number of spatial modes is increased, finer patterns arise in the designs and unit cells tend to be less anisotropic, (D) To generate non-square unit cells, the directions along which the function is periodic can be varied. Rectangular, oblique, and rhombus-shaped unit cells are shown for a fixed unit cell pattern.}	\label{fig: periodic functions}
	\end{center}
\end{figure}

Our goal is to identify the range of effective anisotropic elasticity tensors that can be achieved from periodic single-scale unit-cells composed of two isotropic phases. 
To accomplish this, we use a pixelated representation for the unit cell geometry parametrization.
Exploring all possible combinations of two phases in this pixelated representation is computationally NP-hard\footnote{For example, a 100 $\times$ 100 periodic pixelated representation requires the computation of mechanical properties of $2^{99 \times 99}$ = $2^{9801}$ unit cells.}.
Therefore, to sample periodic diverse anisotropic unit cells efficiently with lower degrees of symmetry, we follow an approach introduced in our prior work \citep{boddapati_single-test_2023}.
This approach is inspired by Cahn's method of generating Gaussian random fields \citep{cahn_phase_1965,soyarslan_3d_2018,kumar_inverse-designed_2020}.
Additionally, this method is inspired by the observation that the power spectral density of microstructures in metallic systems tends to be sparse and has peaks highly concentrated at lower spatial frequencies \citep{zhang_shape-based_2002,yu_characterization_2017}, explained in detail in \cref{sec: Fourier Explanation} and \cref{fig: Fourier Spectrum Comparsion}. 
We first define a periodic function $f(x_1,x_2)$, as a linear combination of cosine functions: 

\begin{equation}
f(x_1,x_2) =  \sum_{m,n} A_{mn}\cos\left(2 \pi(m x_1+n x_2)\right), \quad \forall (x_1,x_2) \in [-0.5,0.5], \quad \forall m,n \in \mathbb{Z}, \label{eq: periodic functions}
\end{equation}
where $m,n$ are spatial frequencies, and $A_{mn}$ are the corresponding cosine function weights \cref{fig: periodic functions}A. 
The function is then thresholded at different values, to generate a family of unit cells, as shown in \cref{fig: periodic functions}B. 

Several notable observations can be made for this approach.
Firstly, the periodicity in the unit cells is ensured by directly selecting cosine functions. 
Interestingly, choosing between sine and cosine as the fundamental periodic function does not affect the distribution of the resulting properties. 
By controlling the symmetries in the weights, we can control the symmetries in the unit cell. 
For instance, if $A_{mn} = A_{nm}$, the result is diagonally symmetric unit cells.
Adjusting the pixel density allows us to generate unit cells with arbitrary resolutions as need-based. 
By increasing the threshold value of the function \cref{eq: periodic functions}, the fill fraction of the stiff phase increases monotonically. 
Each function exhibits a different rate of monotonic increase in the fill fraction, as illustrated in the top inset of \cref{fig: periodic functions}B. 
In this particular work, a few refinements have been made to our previous design approach \citep{boddapati_single-test_2023}.  
The number of spatial modes is a hyperparameter 2$\zeta$+1, with $\zeta$ = $\max(m,n)$. 
In \cref{fig: periodic functions}C, the variation of typical feature sizes is shown when the number of spatial modes is varied. 
Increasing the number of spatial modes leads to smaller feature sizes and increased randomness, resulting in unit cells that tend to be less anisotropic. The shape of the unit cell is also an important factor for achieving anisotropic properties and is a design choice that is often overlooked.
By modifying the directions of periodicity in the proposed periodic function definition \cref{eq: periodic functions}, we can also generate non-square unit cells as shown in \cref{fig: periodic functions}D (explained in detail in \cref{sec: Non-square unitcells}). 
Overall, this method allows us to systematically explore a large property space, identify structures that exhibit the desired anisotropic elasticity tensors and is suitable for estimating theoretical bounds on the anisotropic moduli, as discussed in the next section. 
Further, the coefficients $A_{mn}$ are sampled randomly to generate 2000 different functions and the threshold is varied for 100 increments resulting in a database size of 200000 unit cells {as opposed to only 100 unit cells in our previous work \citep{boddapati_single-test_2023}}. 

Each unit cell's effective material properties are then computed using a numerical homogenization scheme \citep{francfort_homogenization_1986,andreassen_how_2014}.
The homogenization scheme is based on length scale separation and follows a two-scale asymptotic expansion of stress equilibrium equation.
The resulting effective properties are equivalent in the average energy stored in the unit cell for all possible loading conditions.
In computing the effective properties, we fix the pixel resolution at 100. 
For the gray phase, we use a stiffer material DM8530 ($E$ = 1 GPa, $\nu$ = 0.3) and for black phase, we use a softer material Tango Black ($E$ = 0.7 MPa, $\nu$ = 0.49), representative of materials from commercial multi-material Connex 3D printer. 
{The material properties are experimentally determined following the ASTM D638-14 standard test method.}
We follow Voigt notation to describe the homogenized constitutive law, assuming plane-strain condition as
\begin{equation}
\left[\begin{array}{l}
\sigma_{{1}} \\
\sigma_{{2}} \\
\sigma_{{6}}
\end{array}\right]=\left[\begin{array}{lll}
C_{11} & C_{12} & C_{16} \\
C_{12} & C_{22} & C_{26} \\
C_{16} & C_{26} & C_{66} \\
\end{array}\right]\left[\begin{array}{c}
\varepsilon_{{1}} \\
\varepsilon_{{2}} \\
2 \varepsilon_{{6}}
\end{array}\right]. \label{eq: ConstIndicialSmall2}
\end{equation}
where $C_{11}, C_{12}, C_{22}, C_{16}, C_{26}, C_{66}$ are the six independent elasticity tensor parameters in a given reference frame,
$\varepsilon_{1}, \varepsilon_{2}$ are the axial strains, $\varepsilon_{6}$ is the shear strain,  $\sigma_{1}, \sigma_{2}$ are the axial stresses, and $\sigma_{6}$ is the shear stress. 
The homogenized elastic properties of one of the unit cells, normalized with Young's modulus of the stiff phase, are shown in \cref{fig: periodic functions}B (bottom inset). 
In order to satisfy thermodynamic stability, parameters along the diagonal of \cref{eq: ConstIndicialSmall2} are always positive, while the parameters on the off-diagonal could be negative.
$C_{11}$ directly relates the axial stress $\sigma_{1}$ with axial strain $\varepsilon_{1}$, while $C_{16}$ relates the axial stress $\sigma_{1}$ with shear strain $\sigma_{6}$, and so on. 
The off-diagonal moduli $C_{16},C_{26}$ are also known as shear-normal coupling parameters. 
{We aim to determine the maximum and minimum values a single parameter can reach relative to the others, as discussed in the following section. This understanding helps us evaluate the extent of shear-normal coupling induced by anisotropy.}

%%%%%%%%%%%%%%%%%%%%%%%%%%%%%%%%%%%%%%%%%
%%%%%%%%% Results and discussion %%%%%%%%
%%%%%%%%%%%%%%%%%%%%%%%%%%%%%%%%%%%%%%%%%
% \newpage
\section{Data visualizations}
\label{sec: rd}
\subsection{Fill fraction plots}
\begin{figure}[!htb]
    \begin{center}
    \centering 
    \includegraphics[width = 0.95\textwidth]{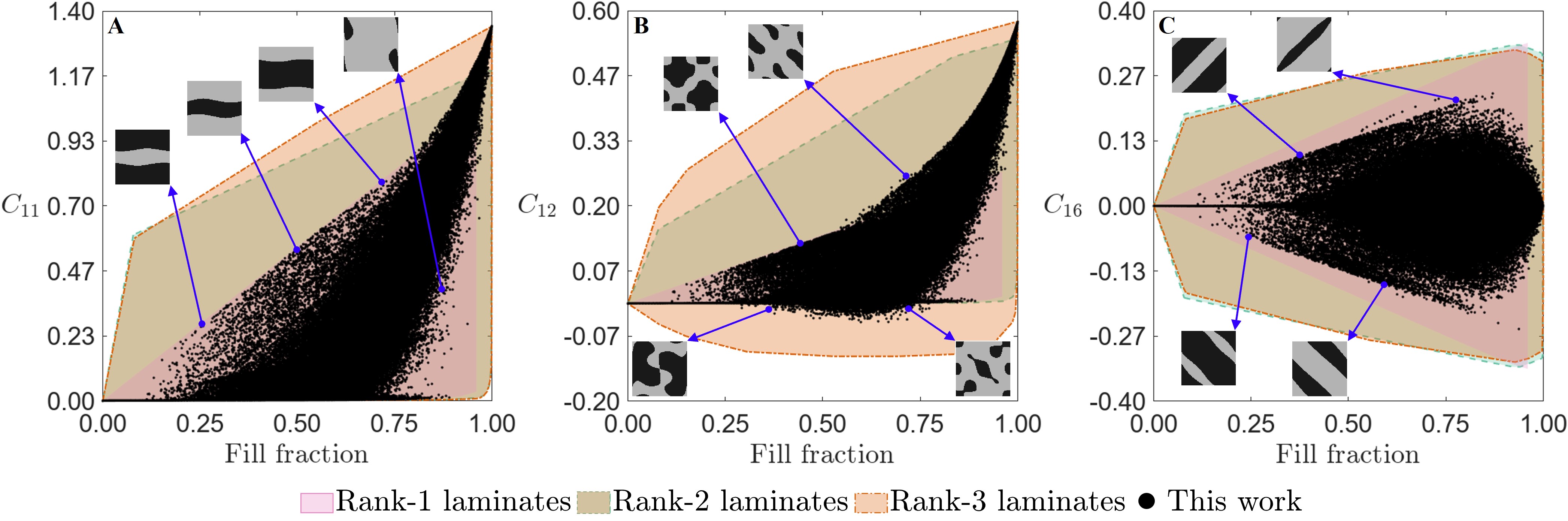}
    \caption{Plots of fill fraction of stiff phase vs. (A)  $C_{11}$, (B) $C_{12}$, (C) $C_{16}$ from this database. All the plots are normalized with the Young's modulus of the stiff phase. Properties of hierarchical laminates are used as theoretical bounds. Rank-1 laminates are indicated with magenta color, rank-2 laminates are indicated with green color and rank-3 laminates are indicated with orange color. Representative unit cells at the boundary are pointed out using arrows. The unit cells away from the bounds contain non-trivial patterns and some of them are displayed in the subsequent figures.}\label{fig: Vol Frac_vs_Ckl}
    \end{center}
\end{figure}

In \cref{fig: Vol Frac_vs_Ckl}, the data of material properties $C_{11}, C_{12}, C_{16}$ is plotted as a function of the fill fraction of the stiff phase. 
All the parameters are normalized with Young's Modulus of the stiff phase.
Note that $C_{22}$, $C_{66}$ have similar property distribution as that of $C_{11}$ while $C_{26}$ property distribution is similar to $C_{16}$ and hence these parameters are not plotted for brevity.
As there are no closed-form expressions for the theoretical bounds on anisotropic moduli, the range of properties achieved by hierarchical laminates upto rank-3 is used as a substitute and indicated in the same plots.
Please refer to \cref{sec: complementary energies} for a discussion on theoretical bounds and \cref{sec: laminates,fig: Hierarchical Laminates} for the construction and computation of the effective properties of the hierarchical laminates.
First, we observe that in all property plots, rank-2 laminates (shown in green) significantly expand the property range compared to rank-1 laminates (shown in magenta). However, the transition from rank-2 to rank-3 laminates (shown in orange) shows minimal improvement across all moduli except for $C_{12}$. 
For $C_{12}$, the negative region is only accessible with rank-3 laminates, and this range shows minimal dependence on the fill fraction beyond 25\%. 
Both rank-2 and rank-3 laminates reduce the strong dependence on achieving higher values for specific moduli, allowing stiffer anisotropic properties beyond linear scaling with a low fraction of the stiff phase. 
Additionally, hierarchical laminates demonstrate that using anisotropic constituent phases can significantly enhance the range of achievable properties in single-scale two-phase composites.

Our database consisted of unit cells achieving the bounds predicted by rank-1 laminates, specifically in the fill fraction ranging between 30\% - 80\%. 
To understand the effect of spatial frequencies, the same plots are plotted in different colors with data constructed from adding a specific number of spatial frequencies (\cref{fig: Vol Frac_vs_Ckl_Spatial}).
It is observed that adding higher spatial modes doesn't necessarily increase the reach in the property space.
In fact, adding higher spatial modes is detrimental to producing anisotropic structures beyond spatial modes of order 3. 
{While square unit cells were sufficient to explore the extremes of the diagonal moduli, the range of the off-diagonal moduli is enhanced with the use of non-square unit cells (\cref{fig: Cij_vs_Ckl_Inclined}). 
For $C_{11}$, the structures at the upper bound are laminate-like structures aligned along the 
$x_1$  direction, while those at the lower bound are laminates aligned along the $x_2$.
For $C_{12}$, the unit cells at the upper and lower bounds are non-laminate-like. In the negative $C_{12}$, the discovered unit cells do not approach the bounds of rank-3 laminates and feature chiral patterns and/or orthogonally aligned thin features. 
For $C_{16}$, the upper bound unit cells are skew laminates tilted towards right, while those at the lower (negative) bound are the same skew laminates, but flipped.}

Hierarchical architectures plays a significant role in structural integrity and functionality across systems of various length scales (e.g. spider silk) \citep{nepal_hierarchically_2023}. 
This hierarchy is often present in non-rectangular coordinate systems. 
For example, wire ropes used in structural engineering contain helical strands arranged in a hierarchical manner, enhancing efficient load transfer and their strength by distributing tensile forces uniformly throughout the cross-section, regardless of the bending direction \citep{feyrer_wire_2007}. 
In biological systems, spinal discs, which are annular cylinders surrounding the spine in the vertebrae, provide shock absorption, support, and flexibility to the spine. 
Their load-bearing protein, collagen, is distributed in a multi-layered laminar fashion \citep{tavakoli_ultrastructural_2018}.
Similarly, the tympanic membrane of the human ear, which is responsible for efficient sound transmission and protecting the delicate structures within the ear, is conical in shape and features collagen structured in a trilaminar fashion with radial and circumferential patterns \citep{volandri_biomechanics_2011}.
These observations along with findings from our work further emphasize the importance of incorporating hierarchical designs to enhance the design capabilities.
\subsection{Pair property plots} 
\label{sec: pair property}
\begin{figure}[!htb]
    \begin{center}
    \centering 
    \includegraphics[width = 0.95\textwidth]{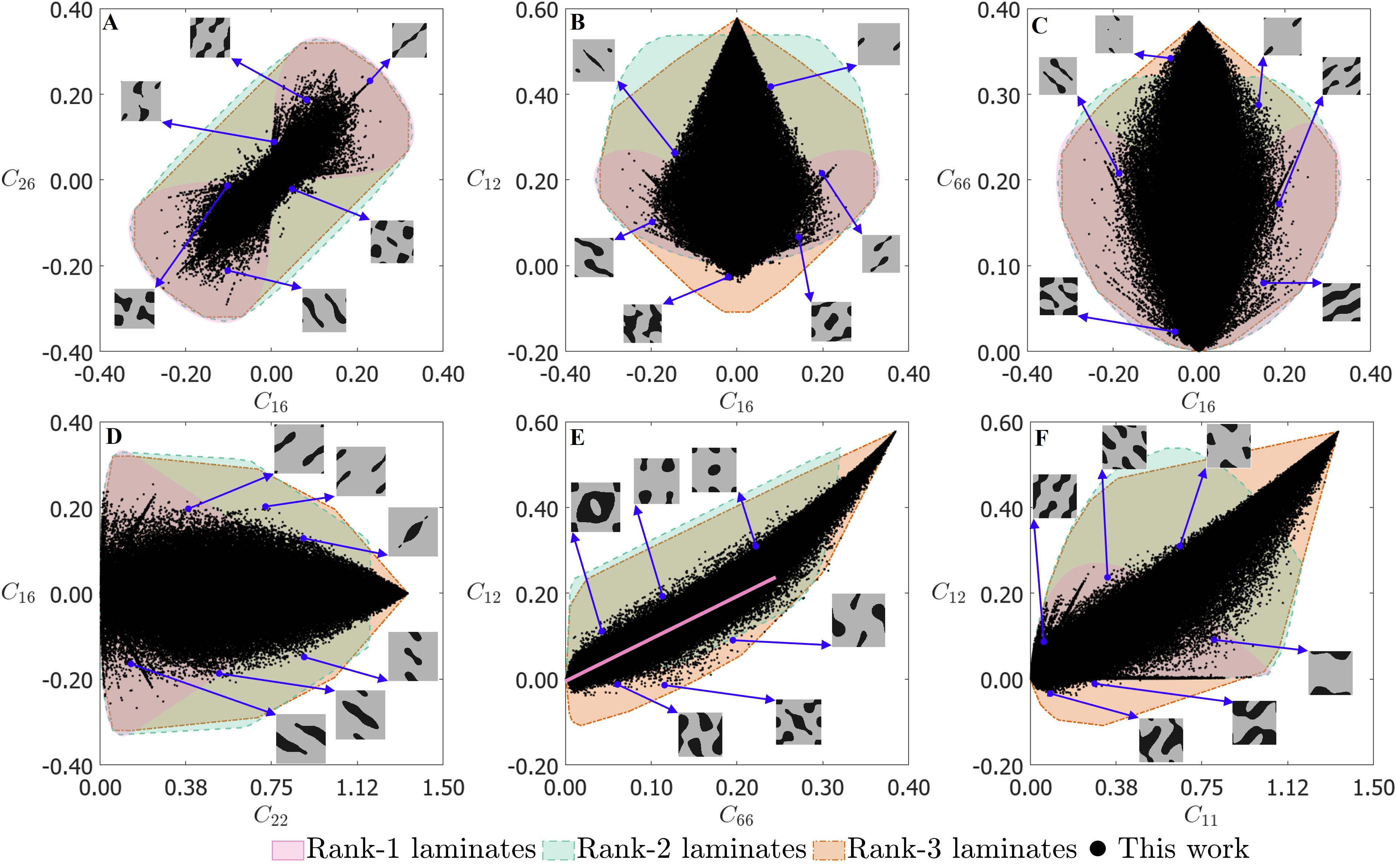}
    \caption{Plots of (A) $C_{16}$ vs. $C_{26}$, (B) $C_{16}$ vs. $C_{12}$, (C) $C_{16}$ vs. $C_{66}$, (D) $C_{22}$ vs. $C_{16}$ (E) $C_{66}$ vs. $C_{12}$ and (F) $C_{11}$ vs. $C_{12}$ from this database. All the plots are normalized with the Young's modulus of the stiff phase shown in gray color. Rank-1 laminates are indicated with magenta color, rank-2 laminates are indicated with green color and rank-3 laminates are indicated with orange color. Representative unit cells at the boundary are pointed out using arrows.}\label{fig: Cij_vs_Ckl}
    \end{center}
\end{figure}

Often, multiple components of the elasticity tensor contribute to the overall mechanical behavior.
Therefore, we also examine the pair property plots. 
For a total of 6 material parameters, there are a total of $\binom{6}{2}$ = 15 distinct pair property plots.
However, due to the symmetry nature of the property plots, it is sufficient to consider only a subset of the plots. 
For example, the property plot of $C_{11}$ vs $C_{16}$ would be the same as $C_{22}$ vs $C_{26}$. 
As the goal of this paper is to identify anisotropic structure-property relations, specifically shear-normal coupling, the property plots associated with the off-diagonal parameters of the elasticity tensor are discussed in detail. 
Therefore, the property plots corresponding to $C_{16}$ vs. $C_{26}$, $C_{16}$ vs. $C_{12}$, $C_{11}$ vs. $C_{22}$, $C_{66}$ . $C_{12}$ are shown in \cref{fig: Cij_vs_Ckl}.
Please refer to \cref{fig: Cij_vs_Ckl_Supp} for the rest of the pair-property plots. 
In the same plots, as discussed earlier, the range of properties achieved by hierarchical laminates are used as a substitute for theoretical bounds (which are plotted separately in \cref{fig: Cij_vs_Ckl_Rk1,fig: Cij_vs_Ckl_Rk2,fig: Cij_vs_Ckl_Rk3} to indicate their distinction).

In the property plot of $C_{16}$ vs. $C_{26}$ shown in \cref{fig: Cij_vs_Ckl}A, we observe that there is a strong correlation along the line inclined at 45$^{\circ}$. 
This means for many of the unit cells, both $C_{16}$ and $C_{26}$ have the same sign. 
Hierarchical laminates uncorrelated this behavior and achieved unit cells with opposing signs for $C_{16}$ and $C_{26}$.
The unit cells identified on the boundary of this property space resemble rank-1 laminates. 
As discussed in \citep{forte_unified_2014}, $C_{16}+C_{26}$ and $C_{16}-C_{26}$ are components of the invariants of the elasticity tensor under coordinate transformation. 
Each of these sums signifies a different contribution to the stored elastic energy (see \cref{eq: I2,eq: J2} in \cref{sec: invariants}).
In the property plots of $C_{66}$ vs. $C_{12}$ and $C_{11}$ vs. $C_{12}$ shown in \cref{fig: Cij_vs_Ckl}E-F, there are generally very few data points with a negative value of $C_{12}$, especially at high values of $C_{66}$.
Again, the negative region for $C_{12}$ is only accessible with rank-3 laminates.
The combination $\kappa = \frac{1}{4}\left(C_{11}+C_{22}+2C_{12} \right)$, known as the bulk modulus, is an invariant. This imposes a restriction on the negative bound of $C_{12}$ to ensure that $\kappa$ remains positive.
The parameter $C_{66} - C_{12}$ is invariant under coordinate transformation. 
Therefore, $C_{66}$ vs. $C_{12}$ plot for rank-1 laminates is strictly a single linear line. $C_{66}$ vs. $C_{12}$ plot for rank-2 and rank-3 laminates is composed of several such lines with different slopes, which is clearly observed in \cref{fig: Cij_vs_Ckl_Rk2}. 

To validate the effective homogenized behavior of unit cells, we conducted experiments on finite tessellations of four different unit cells in a separate study \citep{boddapati_single-test_2023}. 
In this study, we successfully identified all the anisotropic moduli including shear-normal coupling parameters. 
The results showed that at least 10 repeated unit cells in the tessellated structure are required to satisfy homogenization conditions.

% \newpage
\section{Functionally graded structures}
\label{sec: grad}
\subsection{Method of generation of functionally graded structures}
Functionally graded structures have been shown to exhibit unique mechanical behavior such as avoiding shear-banding \citep{chen_functionally_2023,tian_continuous_2024} and, mimicking bone stiffness \citep{kumar_inverse-designed_2020}.
To generate such structures, the parametrization associated with the structure such as truss thickness is often adjusted \citep{panesar_strategies_2018,sanders_optimal_2021,wu_topology_2023}. 
Therefore, these approaches are particularly effective in controlling the isotropic Young's Modulus, relative density, and to some extent, the degree of orthotropic elasticity \citep{li_topology_2018,garner_compatibility_2019}. 
However, achieving smooth spatial gradients in the anisotropic mechanical properties while ensuring the connectivity of adjacent unit cells is challenging.
Here, we illustrate the construction of functionally graded anisotropic structures with seamless transition between unit cells with distinct patterns.

\begin{figure}[!htb]
	\begin{center}
		\centering 
		\includegraphics[width = \textwidth]{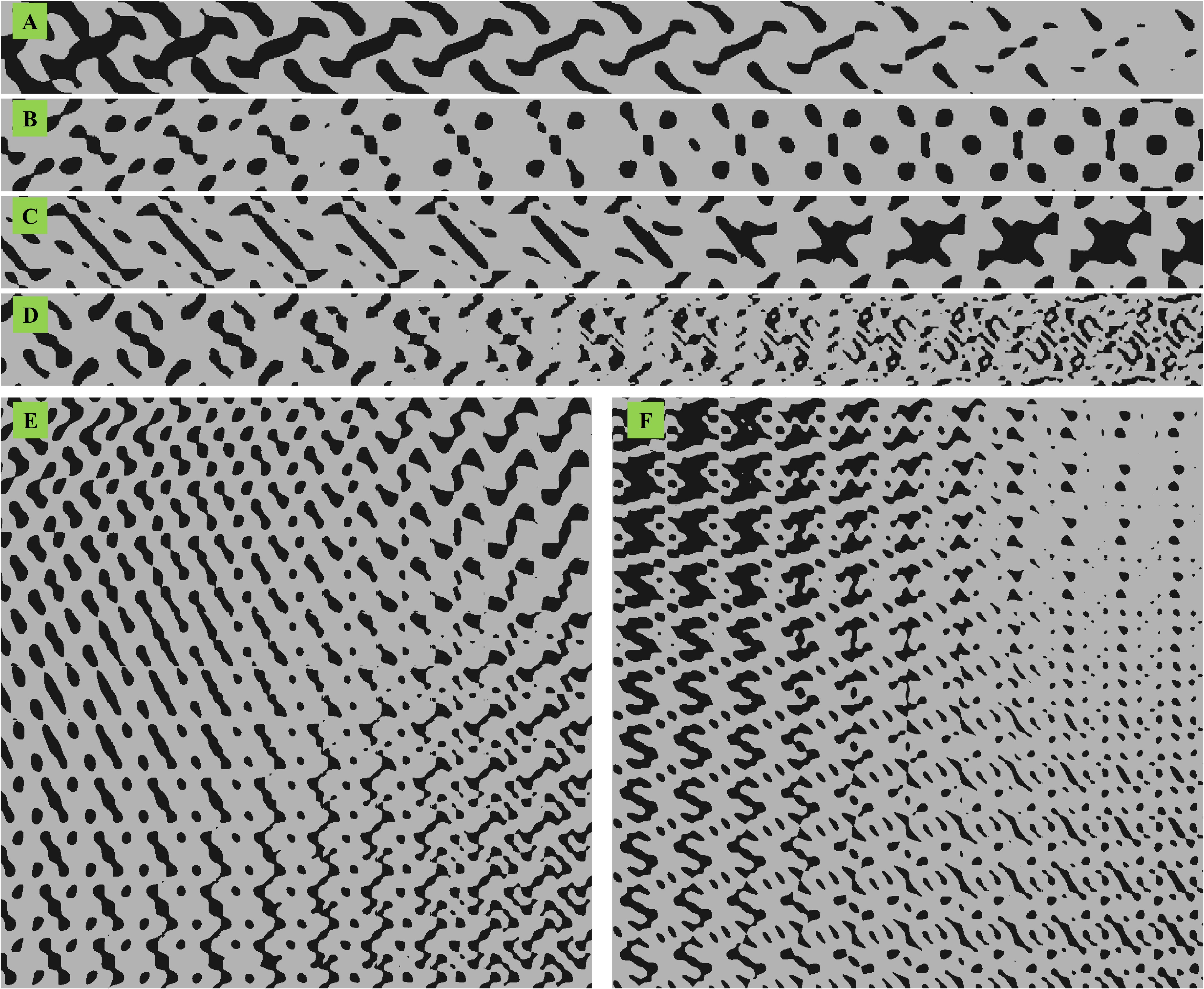}
		\caption{Functionally graded anisotropic metamaterial generation between two unit-cells with different spatial characteristics such as (A) increasing fill fraction of the stiff phase while using the same periodic function, (B) interpolation from asymmetric to symmetric unit-cells by changing the symmetry in the function weights, (C) interpolation between two asymmetric structures with distinct anisotropic properties, (D) interpolation between unit-cells with an increasing number of spatial modes in the periodic function, (E, F) {bilinear} interpolation between four unit cells with different anisotropic behavior at four corners of the boundary.}	\label{fig: gradient structure 1D}
	\end{center}
\end{figure}

\begin{figure}[!htb]
	\begin{center}
		\centering 
		\includegraphics[width = 0.85\textwidth]{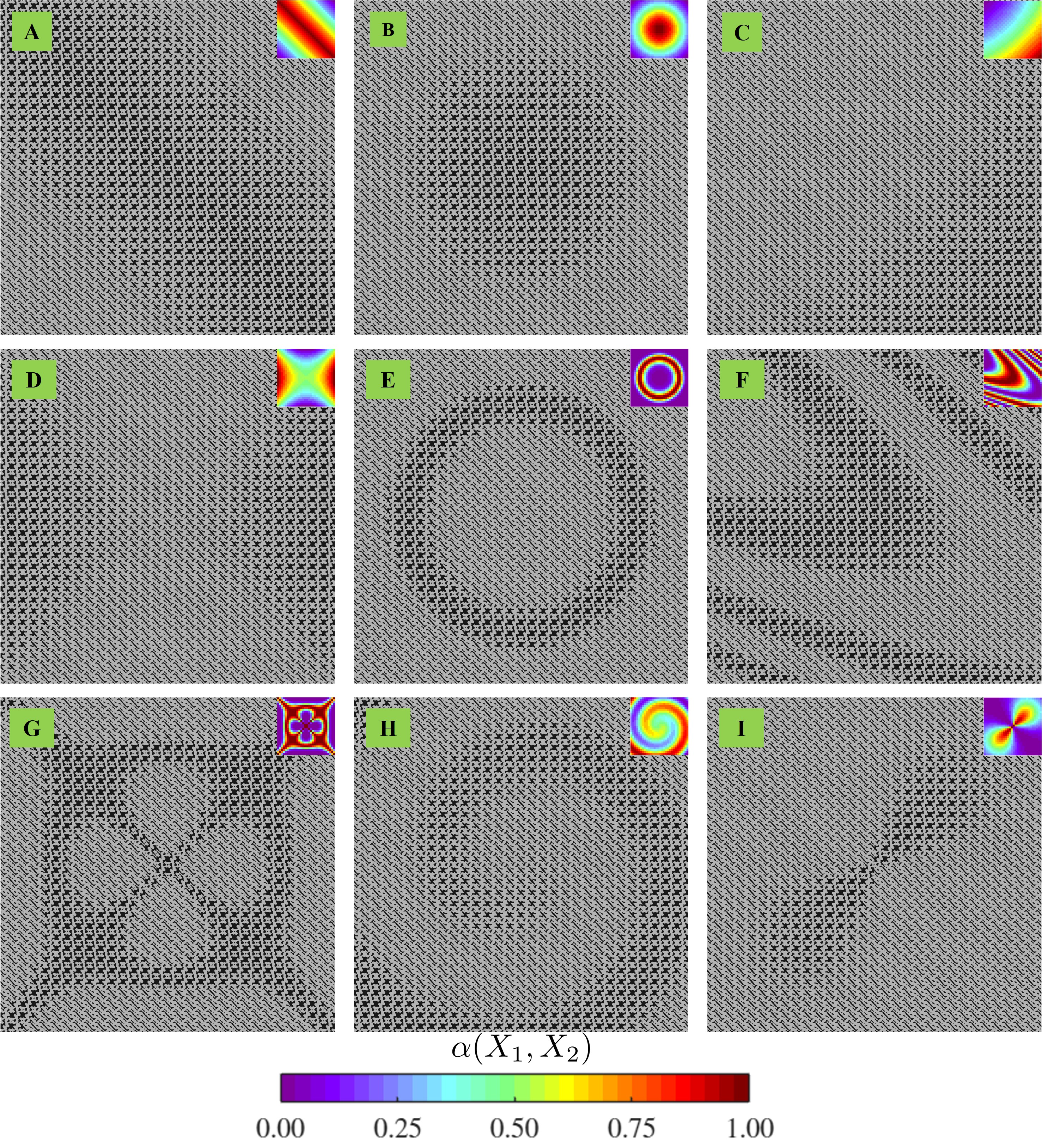}
		\caption{Functionally graded metamaterials with various nonlinear interpolations: (A) diagonal, (B) circular, (C) semi-circular, (D) hyperbolic, (E) annular, (F) parabolic, (G) star, (H) spiral, and (I) orbital. The inset colormap illustrates the variation of the interpolation parameter $\alpha(X_1,X_2)$, transitioning from blue to red, showing how it changes from one unit cell to another. Each graded structure contains 30 $\times$ 30 unit cells. The 1D linear gradient depicted in \cref{fig: gradient structure 1D}C is utilized for all the nonlinear interpolations presented.}	\label{fig: Nonlinear gradient structure}
	\end{center}
\end{figure}

For this purpose, we use the proposed functional representation used to generate unit cells in \cref{sec: database generation}. 
We introduce the local variables $x_1,x_2$ as well as the global variables $X_1,X_2$. 
The global variables are defined only on a coarser grid, while the local variables are defined on a finer grid.
In a graded structure with $p \times q$ unit cells, {with $p,q \in \mathbb{Z}^+$}, the global variables change at discrete locations given by the unit cell centers, while the local variables $X_1,X_2$ change at each pixel location within the unit cell {at a given $x_1,x_2$}. 
The function that is required to generate the graded structure $h(x_1,x_2,X_1,X_2)$ is thus given by
\begin{subequations}
    \begin{align}
    h(x_1,x_2,X_1,X_2) &= \beta(X_1,X_2) f_1(x_1,x_2) + \alpha(X_1,X_2)f_2(x_1,x_2), \label{eq: 1d gradient} \\
    &= \beta(X_1,X_2) \sum_{m,n} A_{mn}\cos\left(2 \pi(m x_1+n x_2)\right) + \alpha(X_1,X_2) \sum_{m,n} B_{mn}\cos\left(2 \pi(m x_1+n x_2)\right), \label{eq: 1d gradient 2}
\end{align}
\end{subequations}

where $\alpha(X_1,X_2)$, $\beta(X_1,X_2) \in [0,1]$ are weighing parameters such that $\alpha(X_1,X_2)+\beta(X_1,X_2)=1$. 
An increase in $\alpha$ signifies the increase in the contribution of second function $f_2(x_1,x_2)$ in the interpolated unit cell. 
The threshold {to generate graded structure from} the function is subsequently set by a bilinear interpolation determined by the thresholds set for the unit cells at the ends. 
This method allows for independent control of several functional gradients, such as porosity, anisotropic moduli, and symmetry. 
In \cref{fig: gradient structure 1D}A-D, various functionally graded structures are shown with different gradients along the $X_1$-axis using \cref{eq: 1d gradient} while maintaining periodicity along the $X_2$-axis.
\cref{fig: gradient structure 1D}E-F shows the bilinear interpolation between four unit cells with different anisotropic behavior at four corners of the boundary. 
By using nonlinear{ly interpolated weighing parameters}, various graded designs such as radial, elliptical, spiral, star, and many more can be created. 
\cref{fig: Nonlinear gradient structure} illustrates the designs interpolated non-linearly using {the unit cells depicted in \cref{fig: gradient structure 1D}C}.
More examples on the graded designs are added to the supplementary information (see \cref{fig: Gradients 1D Fixed Function}, \cref{fig: Gradients 1D Asymmetry to Symmetry}, \cref{fig: Gradients 1D Asymmetry to Asymmetry}, \cref{fig: Gradients Low to high spatial}, \cref{fig: Gradients 2D Linear}). 
In the next section, we investigate the mechanical behavior in two gradient structures with nonlinear interpolations in eliciting atypical mechanical behavior. 

\subsection{Mechanical behavior of graded structures}
\subsubsection{Selective elastic energy localization}
Energy localization refers to the phenomenon where strain energy in a material or structure is concentrated in specific regions. 
Energy localization finds use in applications such as mechanical sensing \citep{lee_acoustic_2023}, and energy harvesting \citep{lee_piezoelectric_2022}.
% , impact mitigation, \CD{crack redirection} 
Here we show that graded structures with anisotropic unit cells achieve selective energy localization, i.e., different localization behavior under different loading conditions.
we achieve this using functionally graded structures with strategically selected unit cells in a radially interpolated design.
The unit cells are chosen such that unit cell \#1 at the boundary is obtained by a 90$^\circ$ rotation of the unit cell \#2 at the interior.
The fill fraction of the stiff phase of both the unit cells is 80\% respectively.
Therefore, this choice makes most of the unit cells in the 20 $\times$ 20 interpolated structure also have uniform fill fractions close to 80\%.
{A uniform fill fraction is selected in the graded design to isolate the impact of differences in stiffness parameters from unit cells that may also be affected by varying fill fractions.}
The (vectorized) stiffness tensor of the unit cell \#1 at the boundary is [0.698  0.131  0.221  0.138  0.095  0.150]$^T$\footnote{In the vectorized format, the stiffness components are ordered as $[C_{11}, C_{12}, C_{22}, C_{16}, C_{26}, C_{66}]^T.$}. 
Therefore, the (vectorized) stiffness tensor of the unit cell \#2 is [0.221  0.131  0.698  0.095 0.138 0.150]$^T$. 

\begin{figure}[!htb]
	\begin{center}
		\centering 
		\includegraphics[width = \textwidth]{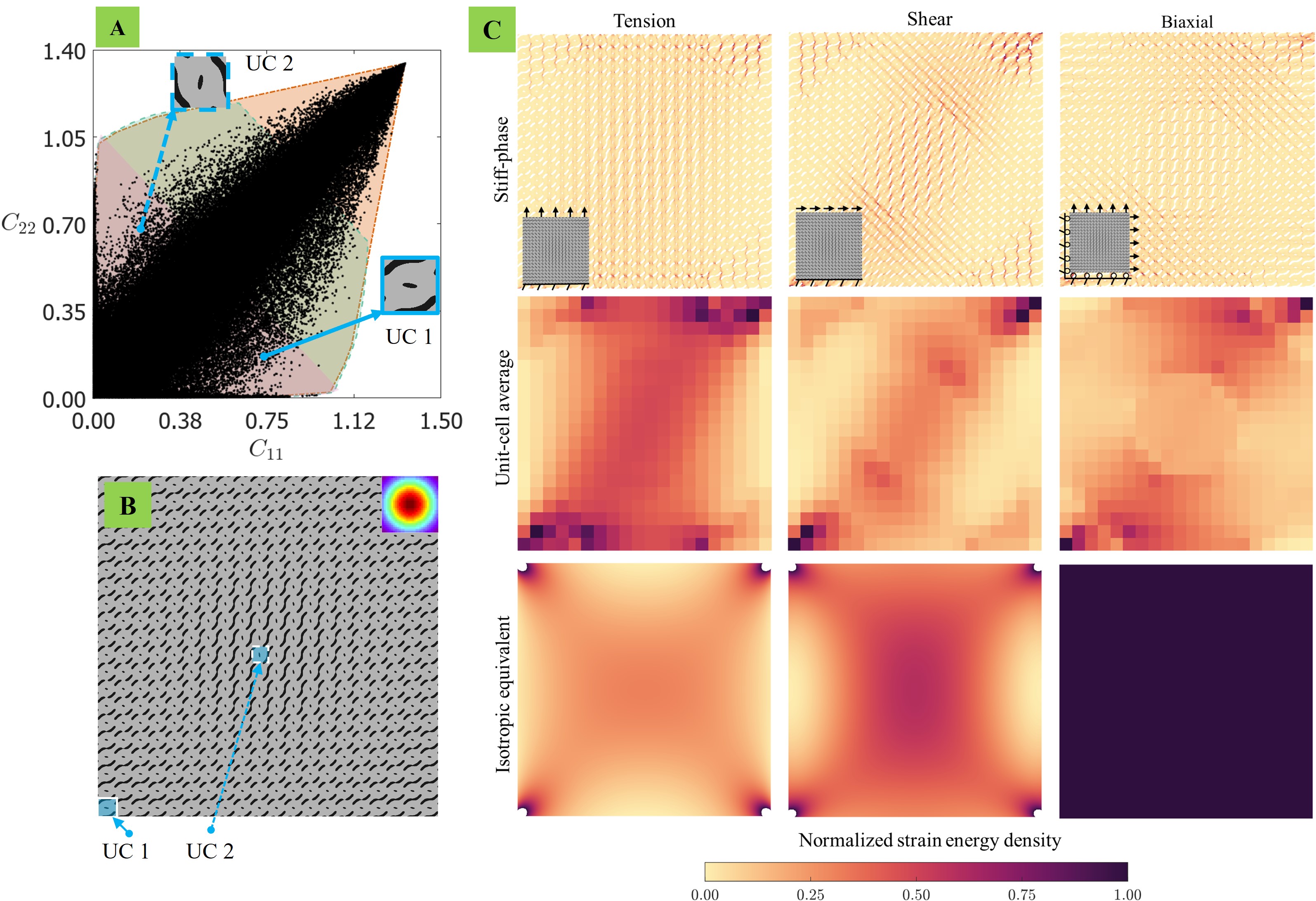}
		\caption{Demonstration of selective energy localization: (A) Unit cell selection based on the extremity in the property space plot of $C_{11}$ vs. $C_{22}$. (B)  Radially graded design with 20 $\times$ 20 tessellation from the chosen unit cells named UC1, UC2. The inset color map shows the variation of interpolation parameter $\alpha(X_1,X_2)$. (C) Distribution of (normalized) elastic energy stored in the circularly interpolated structure for tensile, shear, and biaxial loading displaying selective energy localization arising from anisotropy of the unit cells. The first row shows the energy distribution in just the stiff phase, the second row shows the energy averaged over each unit cell, the third row shows the energy distribution in a continuum-isotropic equivalent.}	\label{fig: Selective Energy Localization}
	\end{center}
\end{figure}

We subject this graded structure to three different loading conditions namely, tension along $x_2$ direction, simple shear along applied on the top edge towards $x_1$ direction, and a biaxial tensile loading by prescribing a displacement of 1.5 mm. 
In \cref{fig: Selective Energy Localization}, the elastic energy density stored in the structure, defined as $W = \frac{1}{2}\bm{\sigma}:\bm{\varepsilon}$ is obtained from finite element analysis (FEA). 
Here we have used the Einstein summation convention, assuming summation over repeated indices, and the double dot ``:'' indicates a double contraction of indices. 
The energy density is plotted first just in the stiff phase, then as areal (volumetric) average at each unit cell while including both the phases.
The energy distribution is then compared with an effective isotropic medium. 
The isotropic equivalent is calculated by replacing unit cell with a material whose bulk and shear moduli are that of Hashin-Shtrikman upper bound for the corresponding fill fraction. 
We observe that under tensile loading the energy distribution becomes significantly localized in a few unit cells in the central region.
In contrast, for the simple-shear boundary condition, the energy is localized in the diagonal region.
For the biaxial loading condition, energy is distributed in the region exterior to the center \footnote{It should be noted that the maximum value of the energy density for three loading cases is different. We are interested in {the distribution over the exact values of the energy density.}}. 
The unit cell \#2 in the interior has a higher $C_{22}$ compared to the unit cell \#1 at the boundary. 
Therefore, under tensile loading along $x_2$ direction, the interior region acts stiffer compared to the exterior region.
This geometric frustration results in increased stresses in the interior region. 
Subsequently, the energy is significantly localized in the center. 
As for shear loading, the shear-normal coupling in the unit cells distributes the stresses along the identified diagonal region.
In the biaxial loading condition, although the fill fraction of all the unit cells is almost same, the interior region is under relatively lower stresses.
If both the unit cells were to be isotropic, this distinctive energy localization is not seen.
The gradient structures can thus localize energy which can be programmed to have a specific failure mode and/or localize stresses and strains in a preprogrammed location.
We further demonstrate the general applicability of this effect when tested with two other unit cells (see \cref{fig: EnergyLocalizationSample2}).

\begin{figure}[!htb]
    \begin{center}
    \centering 
    \includegraphics[width = 0.8\textwidth]{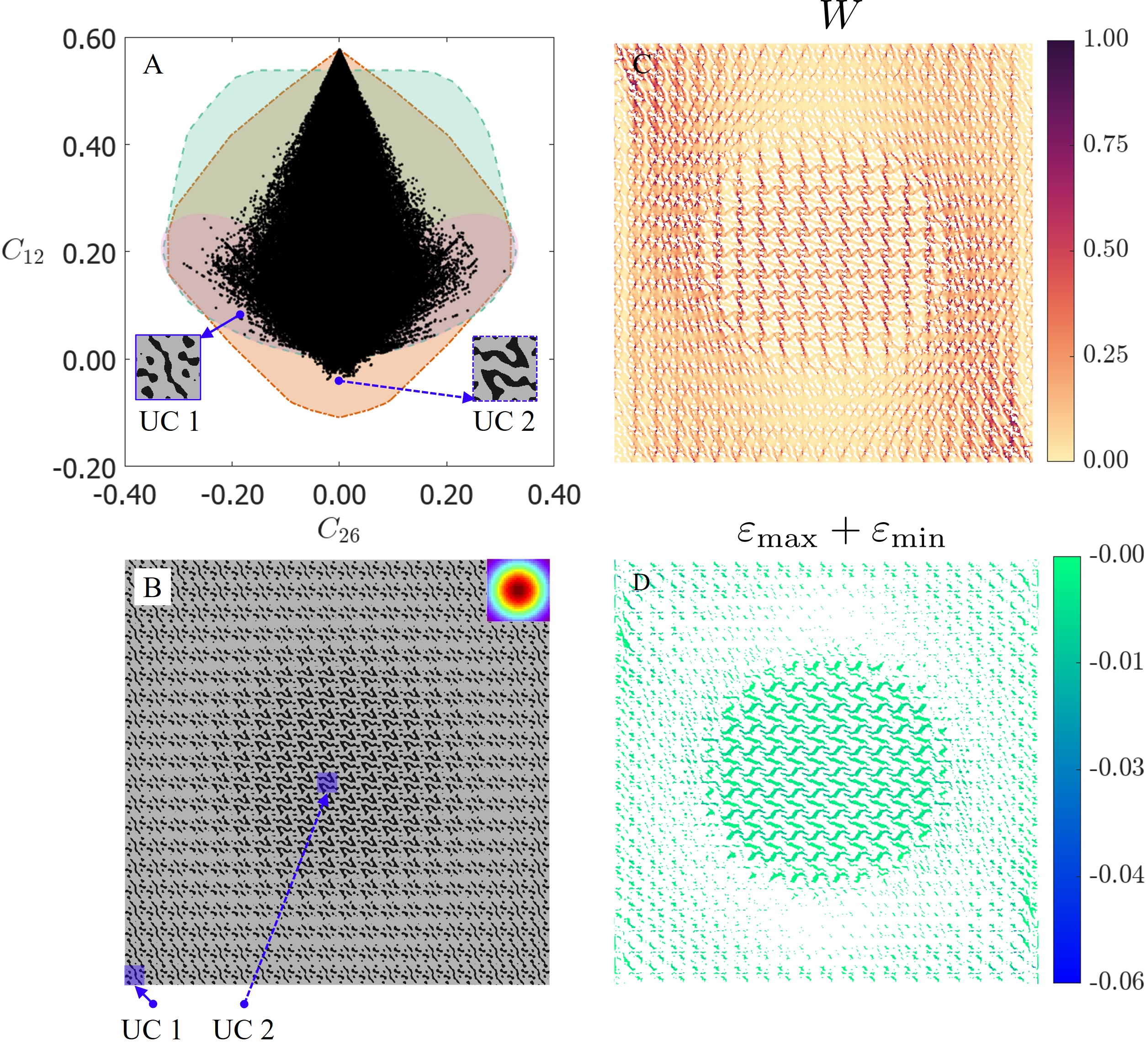}
    \caption{Compressive strains under tensile loading: (A) Unit cell selection based on the extremity in the property space plot of $C_{26}$ vs. $C_{12}$, such that (B) the radially graded design from the selected unit cells that can be additively manufactured using only the stiff phase. (C) Normalized energy distribution in the stiff phase under tensile loading applied along $x_2$ direction. (D) The plot of the sum of principal strains (compressive part only) in the stiff phase which shows compressive strains in the interior region of the radially graded design under the applied tensile loading. Due to geometric incompatibility, the unit cells in the interior undergo compressive strains and compressive stresses under tensile loading.}\label{fig: Energy_Strain}
    \end{center}
\end{figure}

While plotting the energy density distribution provides valuable insights, localization occurs due to all stress and strain components. To delve deeper into this, we select unit cells where the interior region contains those with negative values for 
$C_{12}$, while the exterior region contains positive values for $C_{12}$, as depicted in \cref{fig: Energy_Strain}.
The (vectorized) stiffness tensor of the unit cell \#1 (UC1) at the boundary is [0.134 0.082  0.370  -0.050  -0.120 0.107]$^T$. Similarly, the (vectorized) stiffness tensor of the unit cell \#2 (UC2) in the interior is [0.171  -0.009  0.096  0.001  -0.017  0.248]$^T$. The fill fraction of the stiff phase for the unit cells is [73.7\%, 59.2\%] respectively. The behavior of this design under tensile loading is studied.
In this specific scenario, we observe that unit cells in the interior exhibit chiral characteristics, resulting in lateral expansion akin to auxetic behavior. Conversely, unit cells in the exterior, lacking chirality, contract laterally under tensile loading. 
This geometric mismatch compels interior unit cells to experience compression despite their inherent tendency to expand. 
Consequently, the region with softer-like properties bears greater stresses and strains, leading to a concentration of energy in the center.

\subsubsection{Non-affine deformations}
Non-affine deformations refer to deformations of a material where the local strain or displacement of the material points does not follow the global deformation applied to the material. 
Often soft materials such as polymers, biological tissues, and granular systems exhibit non-affine deformations due to the rearrangement of molecules, particles and/or grains \citep{rubinstein_nonaffine_1997,ellenbroek_non-affine_2009}.
Such non-affine deformations play a crucial role in energy dissipation.
 
Here, we present an example of inducing non-affine deformations in metamaterials on a global scale by utilizing functionally graded structures with strategically selected unit cells.
The two unit cells selected for interpolation are chosen such that their off-diagonal shear-normal coupling moduli are opposite in sign while the other moduli and fill fraction are comparably close. 
We then create an annular interpolated structure as discussed in \cref{fig: Nonlinear gradient structure}C with 20 unit cells along each axis. 
Please refer to \cref{fig: Strain_Localization_Geometries} for more details on the selection of unit cells and their interpolation.
The (vectorized) stiffness tensor of the unit cell in the boundary upon normalization is [0.374, 0.101, 0.108, 0.134, 0.066, 0.110]$^T$.
The (vectorized) stiffness tensor of the unit cell in the annular region upon normalization is [0.115, 0.121, 0.494, -0.084, -0.147, 0.138]$^T$.
The fill fraction of the stiff phase for the unit cells is [62.7\%, 69.5\%] respectively. 
As a result, most of the unit cells in the 20 x 20 interpolated structure have fill fractions close to 65\%.
 
\begin{figure}[!htb]
	\begin{center}
		\centering 
		\includegraphics[width = 0.9\textwidth]{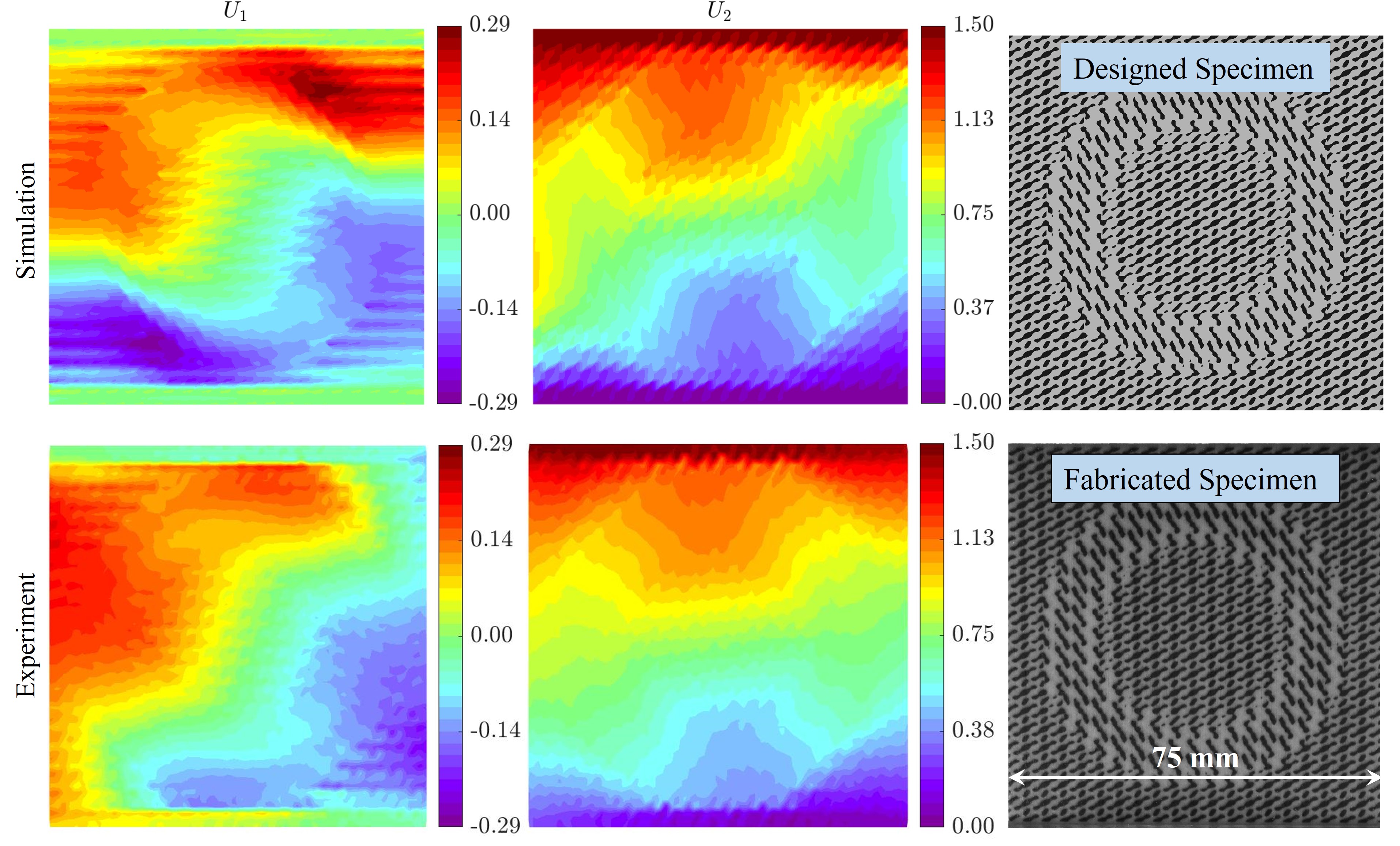}
		\caption{Rotation-like deformation under tensile loading in a gradient structure made with annular interpolation. {The prescribed displacement is 1.5 mm. }The unit cell selection is discussed in \cref{fig: Strain_Localization_Geometries}. The geometric incompatibility between two unit cells with opposing shear-normal coupling behavior leads to non-affine deformation. The top row shows finite element simulation results while the bottom row shows the displacement contours measured using the digital image correlation (DIC) on an additively manufactured specimen.}	\label{fig: Selective Strain Localization 1}
	\end{center}
\end{figure}

This structure is subjected to a tensile loading along $x_2$ direction {by prescribing a displacement of 1.5 mm at the top end}.
In \cref{fig: Selective Strain Localization 1}, the numerical simulations (assuming linear elasticity) reveal that the horizontal displacements exhibit a rotation-like characteristic under this tensile loading.
We further experimentally corroborate the same behavior using full-field measurements obtained using digital image correlation (DIC). More details on the experimental procedure can be found in \cref{sec: exptl data}.
This non-affine deformation behavior seems to arise from the incompatibilities in the off-diagonal shear-normal coupling moduli ($C_{16}, C_{26}$) between the two selected unit cells.
The unit cells with positive shear-normal coupling moduli have a preferential direction to shear under tension, which is opposite to the preferential direction of the unit cells with negative shear-normal coupling moduli.
Therefore this incompatibility in the preferential direction to shear under tension creates internal torque and directs the annular region to rotate while extending in the $x_2$ direction. 
In \cref{fig: Selective Strain Localization 2}, we demonstrate that the observed behavior can also be seen in other pairs of unit cells with opposing shear-normal coupling moduli. 
Additionally, experimentally measured strains are presented for this example in \cref{fig: Selective Strain Localization Strains}, indicating the localization of strain around the annular region.

\begin{figure}[!htb]
	\begin{center}
		\centering 
		\includegraphics[width = \textwidth]{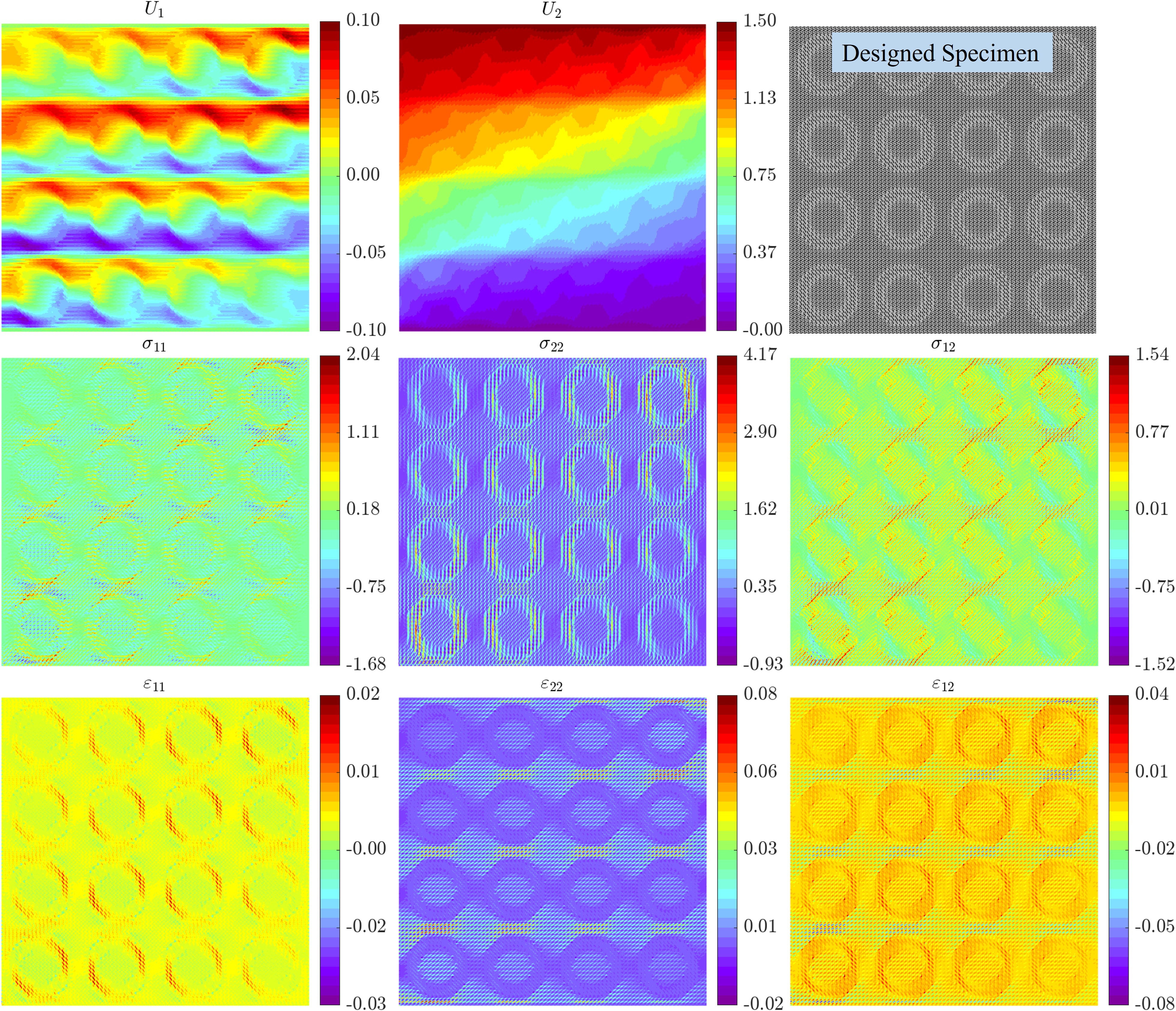}
		\caption{Mechanical behavior in a $ 4\times 4$ {supercell} tessellation design subjected to tensile loading. The {supercell} consists of unit cells with opposing shear-normal coupling arranged in an annular interpolation scheme, which is the entire specimen considered in \cref{fig: Selective Strain Localization 1}. The displacement contour $U_1$ displays multiple regions of rotation-like deformation arising from incompatibilities in the deformation modes of the unit cells. $\sigma_{11},\sigma_{22}$ contours {(in units of MPa)} display how these incompatibilities in {$C_{16}, C_{26}$ lead to alternative regions of compressive and tensile stresses in the tessellated supercell} undergoing tensile loading. $\sigma_{12}$ contour shows the rotation-induced shear stress localization. All the strain contours further corroborate the localization of the strains around the annular interface.}\label{fig: Selective Strain Localization Sample1 4by4}
	\end{center}
\end{figure}

Further, we explore the mechanical behavior of a tessellation obtained by repeating the {supercell} 4 $\times$ 4 times. 
We define the {supercell} as the entire annular interpolated structure shown in \cref{fig: Selective Strain Localization 1} with 20 $\times$ 20 unit cells.
In \cref{fig: Selective Strain Localization Sample1 4by4}, we plot the displacement and strain contours from the FEA on this {supercell} tessellation subjected to tensile loading {by prescribing a displacement of 1.5 mm}  \footnote{Due to computational limitations, we limit our study to a 4 $\times$ 4 tessellation and reduce the unit cell size to 50 pixels from 100 pixels. For this 4 $\times$ 4 {supercell} tessellation, therefore, there are a total of $4\times 4 \times 20 \times 20 = 6400$ unit cells resulting in a finite element mesh with $6400 \times 50 \times 50  = 16,000,000$ elements.}.
We observe that the non-affine rotation-like deformations induced by geometric frustration are present in this {supercell} tessellation at all the annular regions.
Additionally, there is an observed gradient in this non-affine behavior in the horizontal displacement component ($U_1$).
However, the magnitude of the horizontal displacement is reduced compared to the single {supercell}. 
There are non-local interactions from the neighboring annular regions.  
Similar behavior is observed in the case of {supercell} tessellation of example \#2, as shown in \cref{fig: Selective Strain Localization Sample2 4by4}.
This suggests that the length scale {as well as the separation between the annular regions play a significant} role in affecting this rotation-like behavior, indicating the need to utilize micropolar elasticity in the context of such non-affine deformations in mechanical metamaterials \citep{lemkalli_mapping_2023}. 

In the future, a detailed investigation into this scale dependence could be conducted using a multi-scale homogenization approach. 
Further, it is interesting to note that when this structure is subjected to simple shear loading no distinct non-affine deformation is observed, as the unit cells don't differ in the shear-modulus like parameter $C_{66}$.
{Therefore, investigations on the role of incompatibilities in other moduli may unveil new insights into graded structures experiencing geometric frustration under other complex loading conditions.}
Finally, exploring other choices of unit cells with contrasting moduli interpolated non-linearly could reveal unseen atypical mechanical behavior, which we leave for future work.
In metallic materials with microstructure, defects such as grains and grain boundaries serve as strengthening mechanisms by creating incompatibilities that obstruct simple deformation paths.
For example, twin boundaries that arise when a sufficiently high shear load is applied, act as an energy dissipation mechanism contributing to the plasticity of various metals. 
Therefore, the nonlinear interpolations introduced in this work could be utilized to design strengthening mechanisms in metamaterials, {notably for energy dissipation and impact loading,}  extending beyond lattice materials as discussed further in \cite{pham_damage-tolerant_2019}.
 
%%%%%%%%%%%%%%%%%%%%%%%%%%%%%%%%%%%%%%%%%
%%%%%%%%%%%% Conclusion %%%%%%%%%%%%%%%%%
%%%%%%%%%%%%%%%%%%%%%%%%%%%%%%%%%%%%%%%%%
% \newpage
\section{Conclusions}
\label{sec: concl}
In this paper, we estimate the range of anisotropic stiffness tensors achieved by single-scale two-dimensional structured materials.
We compare the property ranges reached by these single-scale architected materials with the extensive property space achieved by hierarchical laminates. 
In all property plots, we observed that rank-2 laminates significantly broaden the property range compared to rank-1 laminates 
We identify regions in the property space, particularly focusing on off-axis shear-normal coupling parameters, where hierarchical designs or the use of two anisotropic constituent phases are necessary to cover a wide property space. 
The bounds estimated alongside the unit cell database could serve as a design tool for the design of extremal metamaterials. 
By utilizing unit cells with extreme anisotropy that lie on the property gamut boundary, we design and fabricate functionally graded metamaterials exhibiting behaviors such as energy localization and localized rotations.
These behaviors are atypical to the corresponding boundary conditions, arising from incompatibilities in the deformation modes of the unit cells.
We then established the concept of supercells which were created by tessellating an entire annular graded structure.
In supercell designs, we observed that the annular regions displayed non-local interactions leading to length-scale dependent behavior. 

Although our study primarily focused on exploring 2D linear elastic properties, the proposed method has the potential to estimate similar bounds for three-dimensional structured materials with shear-shear coupling. 
The diverse range of geometries compiled in our database could also prove useful for studying nonlinear phenomena such as dispersive wave propagation, and viscoelastic behavior in two-phase composites. 
Furthermore, the exploration of other supercell tessellations incorporating different nonlinear interpolation schemes could potentially open up new avenues in the design of multi-scale metamaterials.
% \JB{Need to upload data to metamine website.}

\section*{Acknowledgments}
We would like to thank Prof. Kaushik Bhattacharya (Caltech), Prof. Graeme Milton (University of Utah), and Dr. Andrew Akerson (Caltech) for helpful discussions on the theoretical bounds of the elasticity tensors;
C.D. and J.B. acknowledge the financial support from the US National Science Foundation (NSF) grant number 1835735, {and MURI ARO W911NF-22-2-0109}.

%%%%%%%%%%%%%%%%%%%%%%%%%%%%%%%%%%%%%%%%%
%%%%%%%%%%%%% Appendix   %%%%%%%%%%%%%%%%
%%%%%%%%%%%%%%%%%%%%%%%%%%%%%%%%%%%%%%%%%
% \newpage
\appendix
% \newpage

\section{Database of Elasticity Tensors}
\label{sec: theory}
\subsection{Fourier analysis of pixelated geometries}
\label{sec: Fourier Explanation}
\begin{figure}[!htb]
	\begin{center}
		\centering 
            \includegraphics[width=0.7\textwidth]{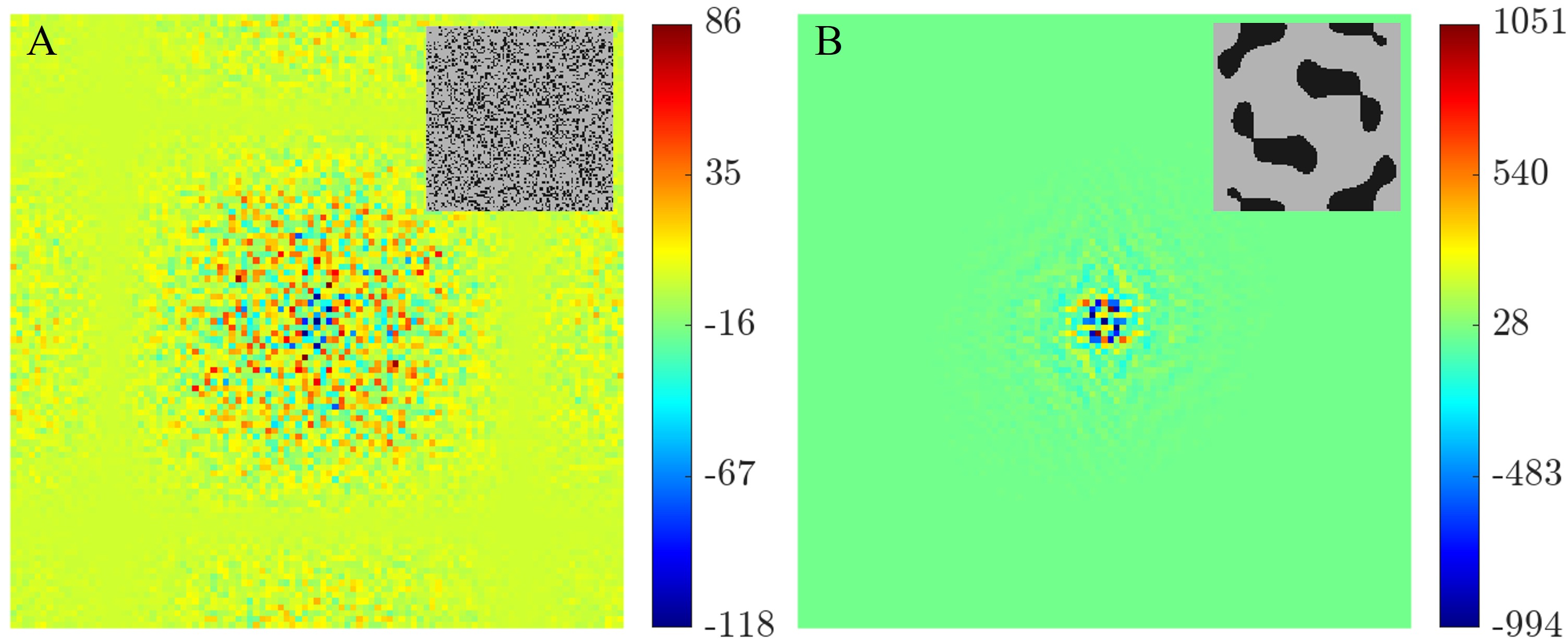}
            \caption{Fourier spectrum comparison of two different type of unit cells: (A) Inset shows a unit cell obtained as a random binary image with it's corresponding real part of the fourier spectrum.  (B) Inset shows a unit cell obtained from the method described in \cref{sec: database generation} with it's corresponding real part of the fourier spectrum. Both the unit cells are chosen such that they have same fill fraction of the stiff phase.}
        \label{fig: Fourier Spectrum Comparsion}
	\end{center}
\end{figure}

Here, we justify the selection of cosine functions for unit cell database generation. Two unit cells are considered for comparison: the first unit cell is randomly generated, lacking distinctive patterns, and appearing almost noisy; the second is obtained using the method in \cref{sec: database generation}. 
After Gaussian filtering, the Fourier spectrum of both unit cells is shown in \cref{fig: Fourier Spectrum Comparsion}. To better illustrate the distribution of frequencies, the zero-frequency component, which represents the image's mean value and has the highest magnitude, is removed from the plots. 
For the almost noisy unit cell, the spectrum shows peaks across a wide range of spatial frequencies. In contrast, the second unit cell's spectrum is concentrated at lower spatial frequencies. This concentration justifies our choice of representation with very small spatial frequencies for generating the unit cells studied in this paper. 

\subsection{Non-square unit cell data generation}
\label{sec: Non-square unitcells}
In a 2D periodic system, the unit cells can be rectangular, square, parallelogram-shaped, or irregularly hexagonal.
Using the concepts of Bravais lattices and Brillouin zone, it is sufficient to consider an arbitrary parallelogram-shaped unit cell defined by the side lengths $a,b$ and the angle between the edges $90^\circ-\theta$, to describe all possible unit cell shapes completely \citep{podesta_symmetry_2019}.
The values for parallelogram $a=1, b=1, \theta = 30^\circ$ would give an equivalent regular hexagonal unit cell.
Therefore, to generate non-square unit cell data, we considered several non-square oblique unit cells when the angle parameter is varied such that $ -45^\circ < \theta <  45^\circ $, while the ratio of side length is varied such that $0.3 \leq \frac{a}{b} \leq 3$.

\subsection{Theory of bounds on anisotropic elasticity tensors}
\label{sec: complementary energies}
For isotropic composites, \cite{hashin_note_1961,hashin_variational_1962} introduced a variational approach to determine the upper and lower bounds on the effective bulk and shear moduli ($\kappa^*$ and $\mu^*$) by decomposing the elastic energy into hydrostatic and deviatoric parts.
However, the elastic energy cannot be decomposed to obtain variational bounds on the independent moduli in the anisotropic case. 
Recently \cite{milton_possible_2017} and \cite{milton_near_2018} addressed this problem in terms of the stress and strain energy pairs and establishing bounds on the sum of the elastic and the complementary energies. 
Let the four energy functions $W_f^k, k=0,1, \ldots, 3$, that characterize the set $GU_f$ of possible elastic tensors $\boldsymbol{C}_*$ be defined by

\begin{subequations}
\begin{align}
W_f^0\left(\boldsymbol{\sigma}_1^0, \boldsymbol{\sigma}_2^0, \boldsymbol{\sigma}_3^0\right) & =\min _{\boldsymbol{C}_* \in G U_f} \sum_{j=1}^3 \boldsymbol{\sigma}_j^0: \boldsymbol{C}_*^{-1} \boldsymbol{\sigma}_j^0, \\
W_f^1\left(\boldsymbol{\sigma}_1^0, \boldsymbol{\sigma}_2^0, \boldsymbol{\varepsilon}_1^0\right) & =\min _{\boldsymbol{C}_* \in G U_f}\left[\boldsymbol{\varepsilon}_1^0: \boldsymbol{C}_* \boldsymbol{\varepsilon}_1^0+\sum_{j=1}^2 \boldsymbol{\sigma}_j^0: \boldsymbol{C}_*^{-1} \boldsymbol{\sigma}_j^0\right], \\
W_f^2\left(\boldsymbol{\sigma}_1^0, \boldsymbol{\varepsilon}_1^0, \boldsymbol{\varepsilon}_2^0\right) & =\min _{\boldsymbol{C}_* \in G U_f}\left[\left(\sum_{i=1}^2 \boldsymbol{\varepsilon}_i^0: \boldsymbol{C}_* \boldsymbol{\varepsilon}_i^0\right)+\boldsymbol{\sigma}_1^0: \boldsymbol{C}_*^{-1} \boldsymbol{\sigma}_1^0\right], \\
W_f^3\left(\boldsymbol{\varepsilon}_1^0, \boldsymbol{\varepsilon}_2^0, \boldsymbol{\varepsilon}_3^0\right) & =\min _{\boldsymbol{C}_* \in G U_f} \sum_{i=1}^3 \boldsymbol{\varepsilon}_i^0: \boldsymbol{C}_* \boldsymbol{\varepsilon}_i^0 .
\end{align}
\end{subequations}

Each energy function $W_f^k, k=0,1, \ldots, 3$, here represents the sum of three elastic energies, each obtained from an experiment where the composite, with effective tensor $\mathbf{C}_*$, is either subject to an applied stress $\boldsymbol{\sigma}_i^0$ or an applied strain $\boldsymbol{\varepsilon}_i^0$.
A total of three stresses $\boldsymbol{\sigma}_i^0$ and $\boldsymbol{\varepsilon}_i^0$ are applied simultaneously on the composite. 
The optimization of these energies to find $\bm{C}_*$ for the applied stresses and strains is non-trivial. 
However, applying the key conclusions from \citep{willis_bounds_1977,avellaneda_optimal_1987,allaire_optimal_1993}, \cite{milton_possible_2017,milton_near_2018} discuss how sequentially layered laminates (or hierarchical laminates) minimize the sum of energies and complementary energies.
In other words, G-closure can be seen as the G-closure of hierarchical laminates which is explained in \cref{fig: Hierarchical Laminates}. 
In two-dimensions, it is sufficient to consider laminates up to rank-3, if the constituent phases are isotropic.
It is also worth noting that hierarchical laminates, with isotropic type effective elasticity tensor, simultaneously achieve Hashin-Shtrikman bulk and shear bounds \citep{francfort_homogenization_1986}. 

\subsection{Construction of hierarchical laminates}
\label{sec: laminates}
\begin{figure}[!htb]
	\begin{center}
		\centering 
            \includegraphics[width=0.7\textwidth]{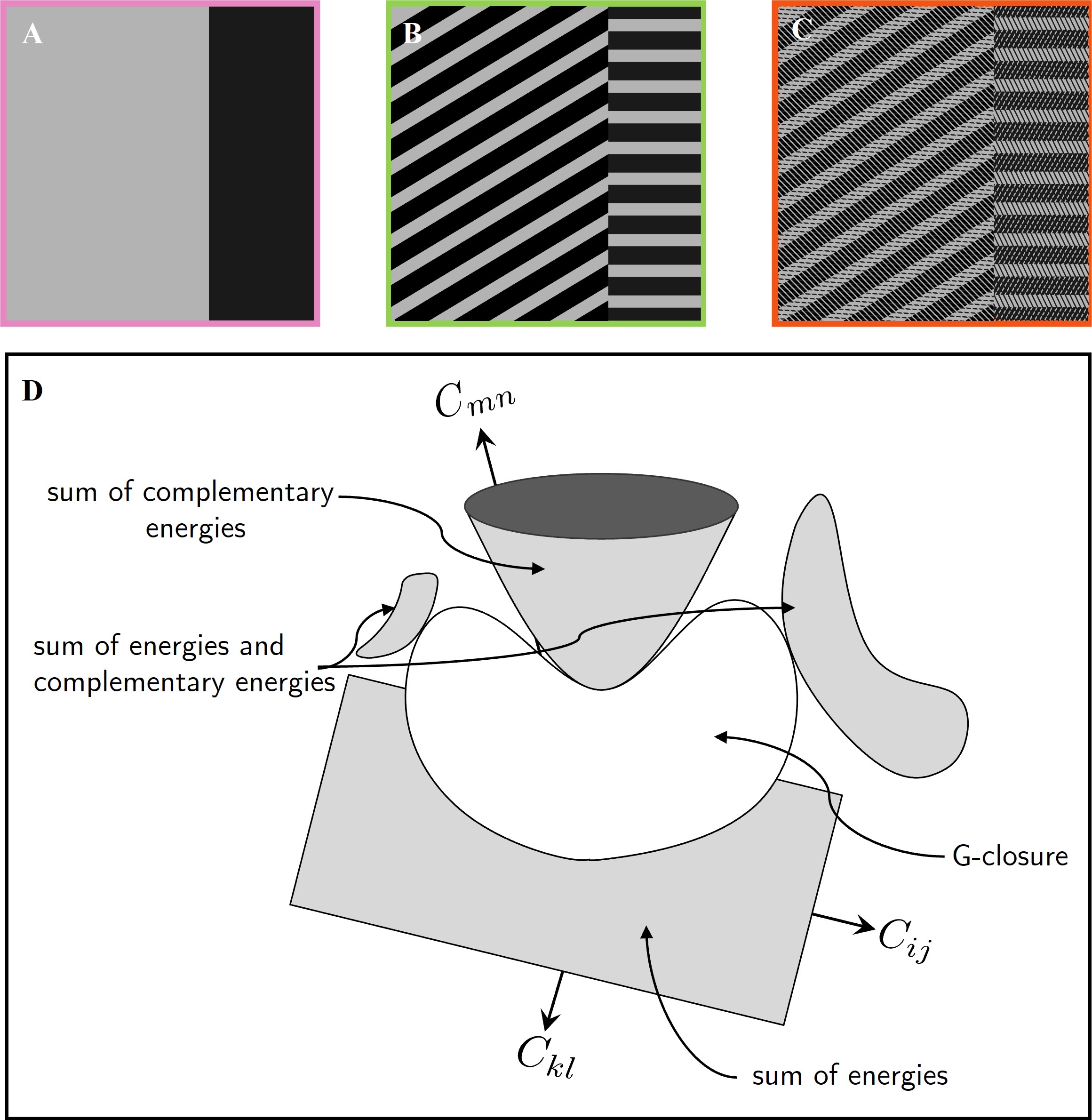} 
            \caption{(A,B,C) Construction of hierarchical laminates of rank-1, rank-2 and rank-3 respectively. For rank-2 laminates, the stiff and soft phases of rank-1 are further laminated in an arbitrarily chosen direction, not necessarily identical to the lamination direction of the rank-1 laminate. Similarly for rank-3 laminates, each rank-1 lamination in rank-2 laminate is laminated again in arbitrary directions. The fill fraction and relative orientation in each sequence of hierarchy are fixed to show the distinction of hierarchy. (D) G-closures are defined by the minimum values of sums of energies and complementary energies. The coordinates represent components of the elasticity tensor. The convexity of the G-closure ensures that the surfaces of energies and complementary energies touch every point tangentially on its boundary (adapted and redrawn from Figure 30.1 in \cite{milton_chapter_2022}). Further, it has been shown that the composites that lie on the boundary of this G-closure are usually hierarchical laminates.}
        \label{fig: Hierarchical Laminates}
	\end{center}
\end{figure}

Multiple-rank laminate materials are hierarchical materials created through an iterative lamination process at increasingly larger length scales. 
A rank-one laminate consists of two isotropic phases, which can be viewed as rank-zero laminates. 
A rank-($m$) laminate is formed by layering a rank-($m-1$) laminate with a laminate of rank-($m-1$) or lower, as shown in \cref{fig: Hierarchical Laminates}.
In two dimensions, it is sufficient to consider laminates up to rank-3 to estimate theoretical bounds, especially when the constituent materials are isotropic, as the 2D elasticity tensor has only three eigenvalues. 
Rank-1 laminates typically have one very high non-zero eigenvalue and two near-zero eigenvalues, while rank-2 laminates have two very high non-zero eigenvalues and one near-zero eigenvalue.
To compute the effective properties of a higher rank-$m$ laminate, the constituent phases are replaced from isotropic to the relevant anisotropic phases of rank-$(m-1)$ laminates (See \cref{fig: Cij_vs_Ckl_Rk1,fig: Cij_vs_Ckl_Rk2,fig: Cij_vs_Ckl_Rk3}). 
A rank-1 laminate is strictly defined by two parameters, the fill fraction of the stiff phase and the angle of orientation of the lamination. 
A rank-2 laminate is defined by four extra parameters the fill fraction of the stiff phase and the angle of orientation of each of the phases of the previous rank-1 laminate.
Similarly, a rank-3 laminate is defined by eight extra parameters.
Suppose, there are 11 different fill fractions and 18 different laminate orientations. 
This results in a total of $11 \times 18 = 198$ elasticity tensors for rank-1 laminates. 
For rank-2 laminates, the total number of possible elasticity tensors is $(11 \times 18)^3 = 7762392$.
For rank-3 laminates, the number increases to $(11 \times 18)^7$. 
First, the effective properties of all the rank-1 laminates are obtained by using two isotropic constituent phases for homogenization.
To compute the effective properties of higher rank-$m$ laminates, the constituent phases are replaced from isotropic to the relevant anisotropic phases of lower rank(See \cref{fig: Cij_vs_Ckl_Rk1,fig: Cij_vs_Ckl_Rk2,fig: Cij_vs_Ckl_Rk3}). 
To reduce the number of computations for the rank-3 laminates database, a subset of randomly chosen rank-2 elasticity tensors (about 1\%) are used. 

%%%%%%%%%%%%%%%%%%%%%%%%%%%%%%%%%%%%%%%%%
%%%%%%%%%%%%% References %%%%%%%%%%%%%%%%
%%%%%%%%%%%%%%%%%%%%%%%%%%%%%%%%%%%%%%%%%
\bibliographystyle{elsarticle-harv}
\bibliography{Bounds_Data,Bounds_Theory}
\newpage

%%%%%%%%Supplementary information%%%%%%%%%
\clearpage
\begin{center}
\textbf{\large Supplementary Information}
\end{center}
%%%%%%%%%% Merge with supplemental materials %%%%%%%%%%
%%%%%%%%%% Prefix a "S" to all equations, figures, tables and reset the counter %%%%%%%%%%
\setcounter{equation}{0}
\setcounter{figure}{0}
\setcounter{table}{0}
\setcounter{page}{1}
\setcounter{section}{0}
\makeatletter
\renewcommand{\theequation}{S\arabic{equation}}
\renewcommand{\thefigure}{S\arabic{figure}}
\renewcommand{\bibnumfmt}[1]{[S#1]}
\renewcommand{\citenumfont}[1]{S#1}
\renewcommand{\thetable}{S\arabic{table}}
\renewcommand{\thesection}{S-\Roman{section}}
%%%%%%%%%% Prefix a "S" to all equations, figures, tables and reset the counter %%%%%%%%%%

%%%%%%%%%%%%%%%%%%%%%%%%%%%%%%%%%%%%%%%%%
%%%%%%%%%%%%% Theory %%%%%%%%%%%%%%%%%%%%
%%%%%%%%%%%%%%%%%%%%%%%%%%%%%%%%%%%%%%%%%
\section{Elasticity tensor invariants}
\label{sec: invariants}
Using harmonic decomposition \citep{forte_unified_2014}, the strain energy density ($W$) can be decomposed as, 
\begin{subequations}
\begin{align}
    2 W&=\sigma_{i j}^{s} \varepsilon_{i j}^{s}+\sigma_{i j}^{d} \varepsilon_{i j}^{d},\\
        &=D_{i j k l} \varepsilon_{i j}^{d} \varepsilon_{k l}^{d}+2 \mu \varepsilon_{pq}^{d} \varepsilon_{p q}^{d}+a_{pq} \varepsilon_{p q}^{d} \varepsilon_{rr}+\kappa \varepsilon_{p p} \varepsilon_{q q}, \label{eq: energy decomposition} 
\end{align}
\end{subequations}

where $\sigma_{i j}^{s}$, $\sigma_{i j}^{d}$, $\varepsilon_{i j}^{s}$,$\varepsilon_{i j}^{d}$ are the spherical and deviatoric part of the stress and strain tensors respectively. 
Here Einstein's notation of summation over repeated indices is used. $\sigma_{i j}^{s}$, $\sigma_{i j}^{d}$ are defined as 
\begin{subequations}
\begin{align}
\sigma_{i j}^{s}&=\left(\frac{1}{2} a_{p q} \varepsilon_{p q}^{d}+k \varepsilon_{p p}\right) \delta_{i j}, \\
\sigma_{i j}^{d}&=D_{i j k l} \varepsilon_{k l}^{d}+2 \mu \varepsilon_{i j}+\frac{1}{2} \varepsilon_{p p} a_{i j}
\end{align}
\end{subequations} 
where $D_{i j k l}$, $a_{i j}$, $\lambda, \mu$ are the parameters obtained from harmonic decomposition of the elasticity tensor as
\begin{equation}
C_{i j k l}= D_{i j k l}+\frac{1}{6}\left(\delta_{i j} a_{k l}+\delta_{k l} a_{i j}+\delta_{i k} a_{j l}+\delta_{j l} a_{i k}+\delta_{i l} a_{j k}+\delta_{j k} a_{i l}\right) +\lambda \delta_{i j} \delta_{k l}+\mu\left(\delta_{i k} \delta_{j l}+\delta_{i l} \delta_{j k}\right), \label{eq: harmonic decomposition}
\end{equation}
where $D_{i j k l}$, $a_{i j}$, $\lambda, \mu$ are all linear combinations of the elasticity tensor parameters.  

 $\lambda$ is defined as
\begin{align}
    \lambda&=\frac{1}{8}\left(3 C_{ {ppqq }}-2 C_{{pqpq}}\right) \quad  \text{with} \quad p, q \in[1,2], \notag\\
&=\frac{1}{8}\left(3\left(C_{1111}+C_{1122}+C_{2211}+C_{2222}\right)-2\left(C_{1111}+C_{1212}+C_{2121}+C_{2222}\right)\right), \notag\\
&=\frac{1}{8}\left(C_{11}+C_{22}+6 C_{12}-4 C_{66}\right). 
\end{align}

 $\mu$ is defined as
\begin{align}
    \mu&=\frac{1}{8}\left(2 C_{p q p q}-C_{p p q q}\right), \notag\\
&=\frac{1}{8}\left(2\left(C_{1111}+C_{1212}+C_{2121}+C_{2222}\right)-\left(C_{1111}+C_{1122}+C_{2211}+C_{2222}\right)\right), \notag \\
&=\frac{1}{8}\left(C_{11}+C_{22}+4 C_{66}-2 C_{12}\right).
\end{align}

$a_{ij}$ is defined as
\begin{align}
a_{ij}=& \frac{1}{12}\left(2 C_{i p j p}-C_{p q p q} \delta_{i j}\right), \notag \\
a_{11} =& - a_{22} = \frac{1}{12}\left(2\left(C_{1111}+C_{1212}\right)-\left(C_{1111}+C_{1212}+C_{2121}+C_{2222}\right)\right. = \frac{1}{12}\left(C_{11}-C_{22}\right), \notag \\
a_{12}=& a_{21} =\frac{1}{12}\left(2\left(C_{1121}+C_{1222}\right)\right) =\frac{1}{12}\left(2\left(C_{16}+C_{26}\right)\right)=\frac{1}{6}\left(C_{16}+C_{26}\right). 
\end{align}

$D_{ijkl}$ is defined as
\begin{align}
D_{1111}=& \frac{1}{8}\left(C_{11}+C_{22}-2 C_{12}-4C_{66}\right), \notag \\
D_{1112}=& \frac{1}{2}\left(C_{16}-C_{26}\right), \notag \\
D_{1111}=& -D_{2211} = -D_{1221} = -D_{2121}= -D_{1212} = -D_{2112} = -D_{1122} = D_{2222},\\
D_{1211}=& D_{2111} = D_{1121} = -D_{2221}= D_{1112} = -D_{2212} = -D_{1222} = -D_{2122}.
\end{align}

Based on this decomposition, four invariants, that do not depend on the choice of the coordinate system of the material, $I_1, J_1, I_2, J_2$ are defined as follows. 

$I_1$ is a measure of the sphericity of the material and is defined as
\begin{align}
    I_1 = \kappa&=\lambda+\mu, \notag \\
&=\frac{1}{4}\left(C_{11}+C_{22}+2C_{12}\right).
\end{align}

$J_1$ is a measure of the isotropic part of the deviatoricity of the material and is defined as
\begin{align}
    J_1 = \mu &=\frac{1}{8}\left(C_{11}+C_{22}+4 C_{66}-2 C_{12}\right).
\end{align}

$I_2$ is the square of a norm that measures the amount of coupling energy in the material and is defined as
\begin{align}
    I_2 &= \sqrt{a_{ij}a_{ij}}, \notag\\
&= \sqrt{2(a_{11}^2+a_{12}^2)}, \notag\\
&= \frac{1}{6\sqrt{2}}\sqrt{(C_{11}-C_{22})^2+4(C_{16}+C_{26})^2} . \label{eq: I2}
\end{align}

$J_2$ measures the squared norm of the anisotropic part of the deviatoricity of the material and is defined as
\begin{align}
    J_2 &= \sqrt{D_{ijkl}D_{ijkl}}, \notag\\
&= \frac{1}{2\sqrt{2}}\sqrt{(C_{11}+C_{22}-2C_{12}-4C_{66})^2+16(C_{16}-C_{26})^2}. \label{eq: J2}
\end{align}

In the case of isotropic material, the invariants $I_2, J_2$ would be zero. 
A few insights on the trends in the property plots, as discussed in \cref{sec: pair property} are derived from the invariants.
% \newpage
\section{Experimental data acquisition}
\label{sec: exptl data}
\textit{Fabrication}: We fabricate the specimens using a commercial multi-material polyjet technology-based 3D printer, the Stratasys Objet500 Connex. 
The specimens measure 75 × 75 × 5 mm, not including the portion inserted into the grips. 
For the stiff phase, we use Stratasys' proprietary material DM8530, and for the soft phase, TangoBlack.
The material properties are experimentally determined following the ASTM D638-14 standard test method: DM8530 has a Young’s modulus (\textit{E}) of 1000 $\pm$ 90 MPa and a Poisson’s ratio ($\nu$) of 0.35, while TangoBlack has a Young’s modulus (\textit{E}) of 0.7 MPa and a Poisson’s ratio ($\nu$) of 0.49. 

\textit{Tension testing}: We subject the additively manufactured specimens to displacement-controlled tension tests using a universal testing machine, Instron E3000.
We apply a vertical displacement of 1.5 mm at the top boundary at a rate of 0.5 mm/min resulting in a global axial strain of $\tilde{\varepsilon}_{22}=0.02$ and a global strain rate of $ 1.1 \times 10^{-4}$ s$^{-1}$. 
Custom-designed grips are fabricated out of aluminum and are serrated to hold the specimens firmly and prevent any lateral slipping.
We use the same strain rate while measuring the constitutive material properties of the individual phases. 
We repeated the experiments on three different specimens for the same design. 
All the specimens showed little variation in the observed global behavior. 

\textit{Digital image correlation}: We use DIC, an image-based optical technique, to measure the full-field displacements \citep{schreier_image_2009, hild_comparison_2012}. 
We capture images at a frequency of 1 Hz using a Nikon D750 camera equipped with a Nikon AF-S NIKKOR 24-120mm f/4G ED VR zoom lens.
We use manual mode at an exposure rate of 1/640 sec, an ISO setting of 1250 and an aperture setting of F8. 
We use the \textit{global DIC} approach, and perform DIC using piece-wise linear shape functions defined on a triangular mesh with an edge length of $\approx$ 0.35 mm, to compute the displacements (as in \cite{agnelli_systematic_2021}). 
%%%%%%%%%%%%
% \newpage
\section{Additional figures}

\begin{figure}[!htb]
    \begin{center}
    \centering 
    \includegraphics[width = 0.95\textwidth]{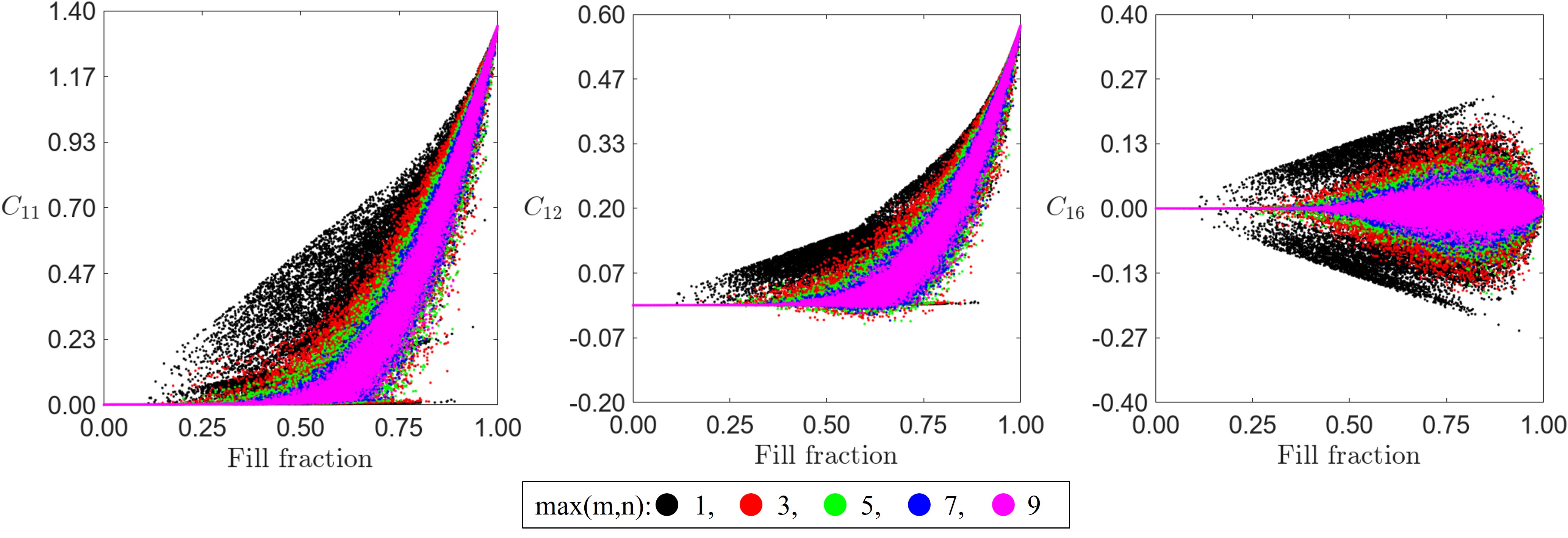}
    \caption{Plots of fill fraction of stiff phase vs. $C_{11}$, $C_{12}$, $C_{16}$ from this database as the maximum number of spatial modes is varied. All the plots are normalized with the Young's modulus of the stiff material. Adding higher spatial modes doesn't increase the span of the material properties.Lower spatial modes exhibit limited shape diversity, while higher spatial modes tend to be isotropic. There exists a balanced middle ground between these two extremes. }\label{fig: Vol Frac_vs_Ckl_Spatial}
    \end{center}
\end{figure}

\begin{figure}[!htb]
    \begin{center}
    \centering 
    \includegraphics[width = 0.95\textwidth]{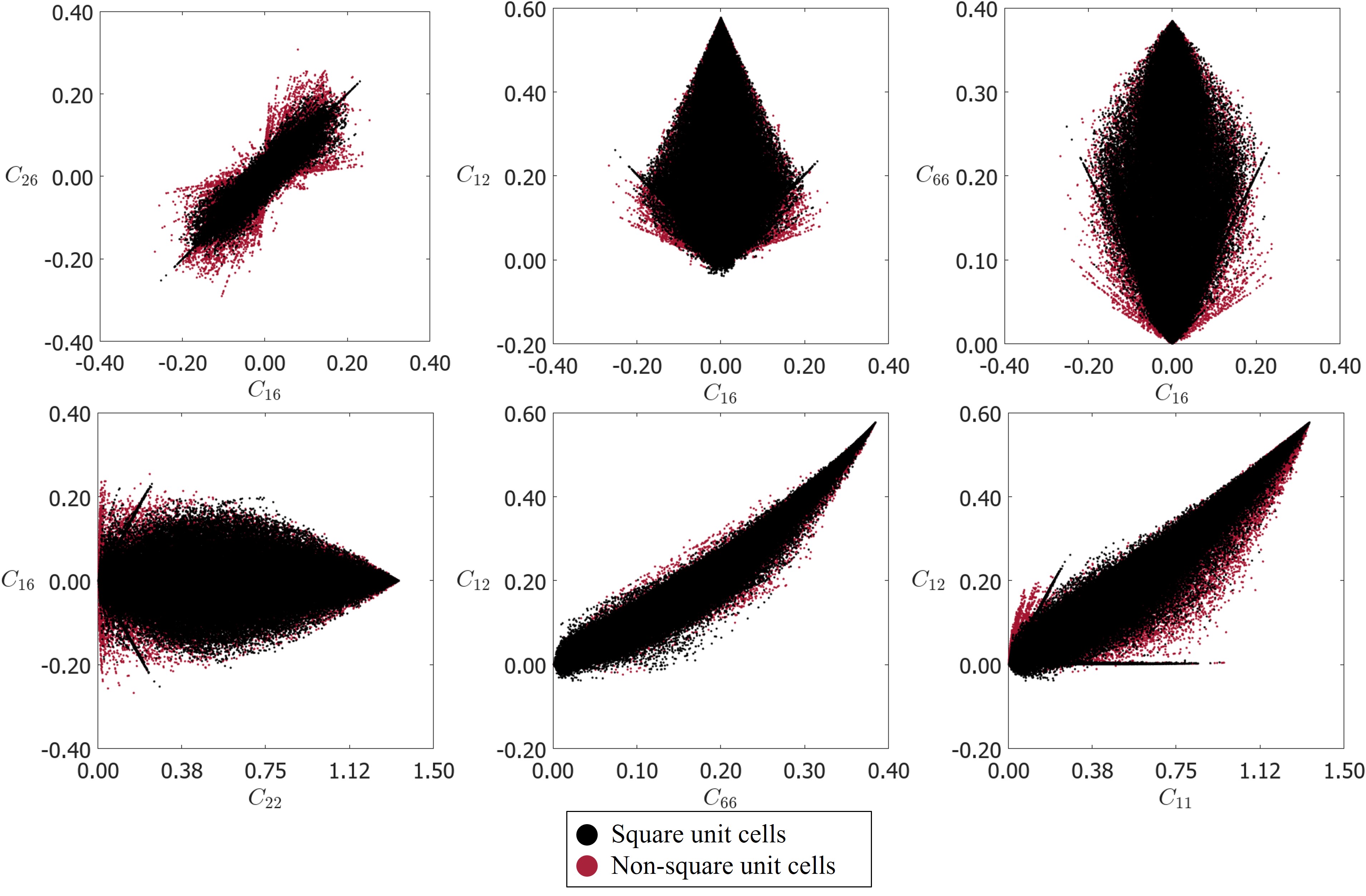}
    \caption{Comparison of the properties between square and non-square unit cells. Non-square unit cells are parallelogram-like shaped and contain patterns that are periodic along non-orthogonal directions. The span in the off-diagonal properties is enhanced with non-square unit cells, especially $C_{16},C_{26}$. Off-diagonal moduli $C_{16},C_{26}$ have high dependence on the unit cell shape and symmetries.}\label{fig: Cij_vs_Ckl_Inclined}
    \end{center}
\end{figure}

\begin{figure}[!htb]
    \begin{center}
    \centering 
    \includegraphics[width = 0.95\textwidth]{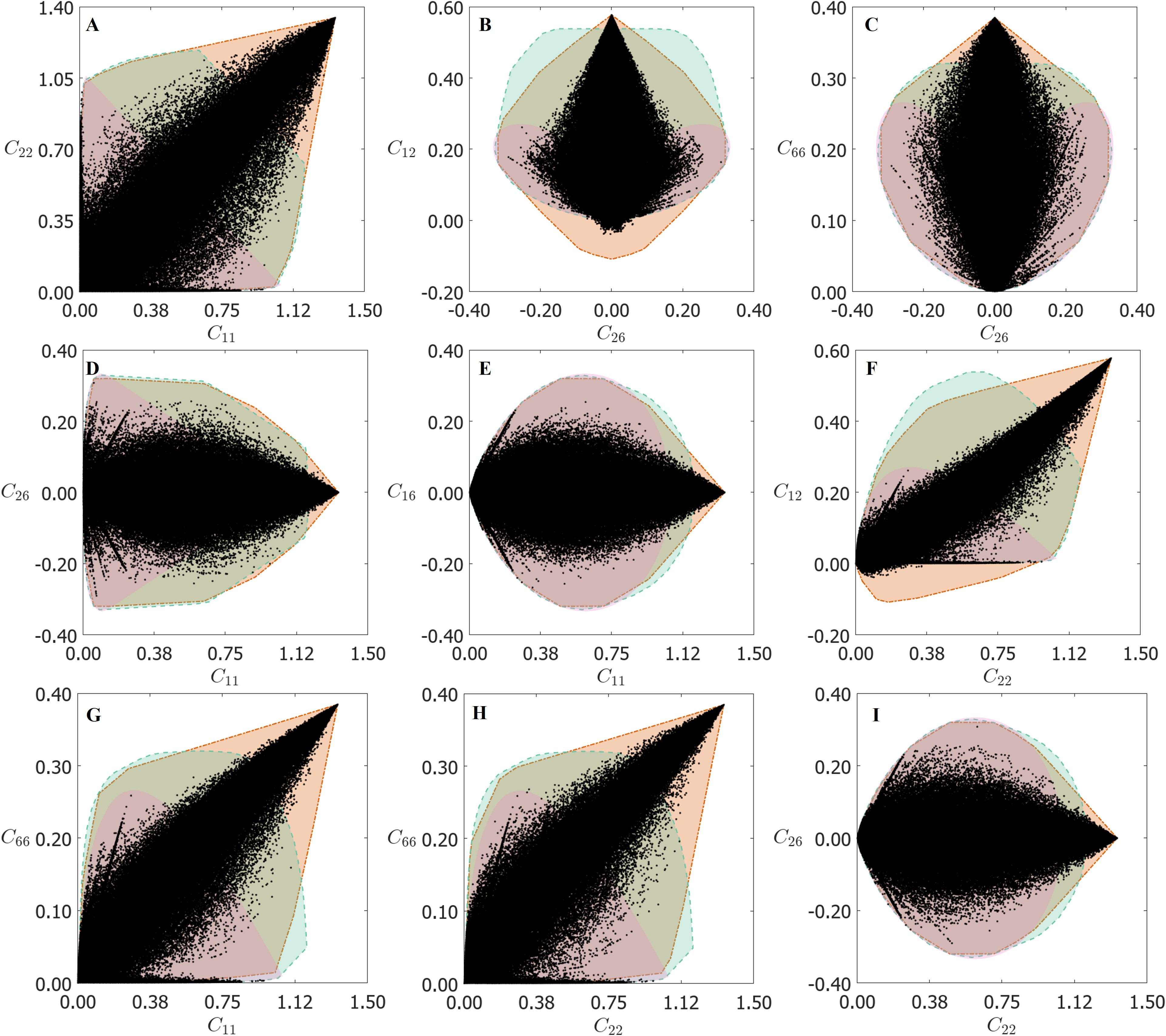}
    \caption{Plots of (A)  $C_{11}$ vs. $C_{22}$, (B) $C_{26}$ vs. $C_{12}$, (C) $C_{26}$ vs. $C_{66}$, (D) $C_{11}$ vs. $C_{26}$, (E) $C_{11}$ vs. $C_{16}$, (F) $C_{22}$ vs. $C_{12}$, (G) $C_{11}$ vs. $C_{66}$, (H) $C_{22}$ vs. $C_{66}$ and (I) $C_{22}$ vs. $C_{26}$ from this database. All the plots are normalized with the Young's modulus of the stiff material. Rank-1 laminates are indicated with magenta color, rank-2 laminates are indicated with green color and rank-3 laminates are indicated with orange color.}\label{fig: Cij_vs_Ckl_Supp}
    \end{center}
\end{figure}

\begin{figure}[!htb]
    \begin{center}
    \centering 
    \includegraphics[width = 0.8\textwidth]{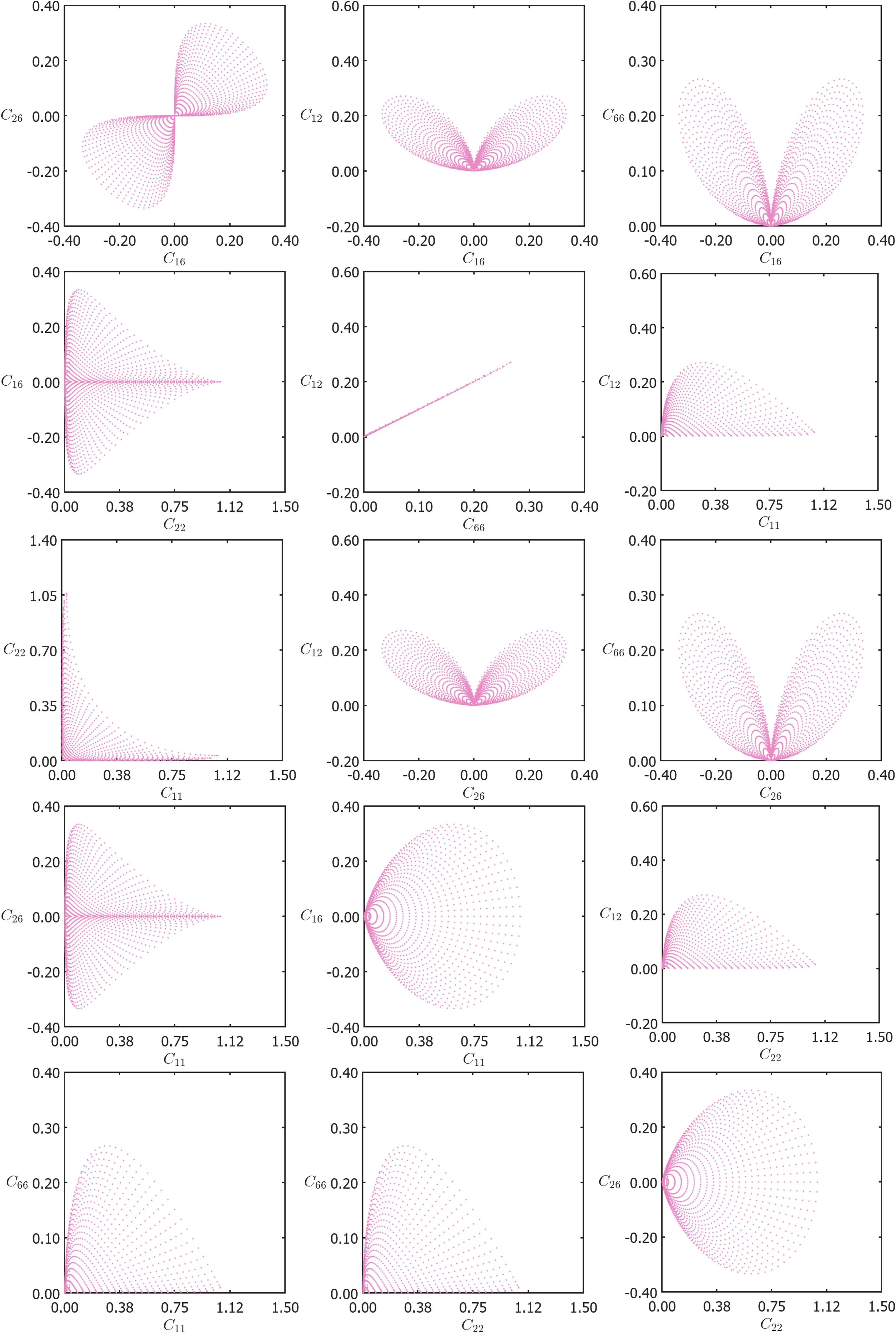}
    \caption{Pair property plots for rank-1 laminates. First, the properties of the laminates whose normal direction is aligned with $x_1$ direction are computed, as the fill fraction of stiff phase is incremented in intervals of 0.04 up to 0.96. Subsequently, a coordinate transformation is applied to these tensors to determine the properties of laminates whose normal is not aligned with the coordinate axes.}\label{fig: Cij_vs_Ckl_Rk1}
    \end{center}
\end{figure}

\begin{figure}[!htb]
    \begin{center}
    \centering 
    \includegraphics[width = 0.8\textwidth]{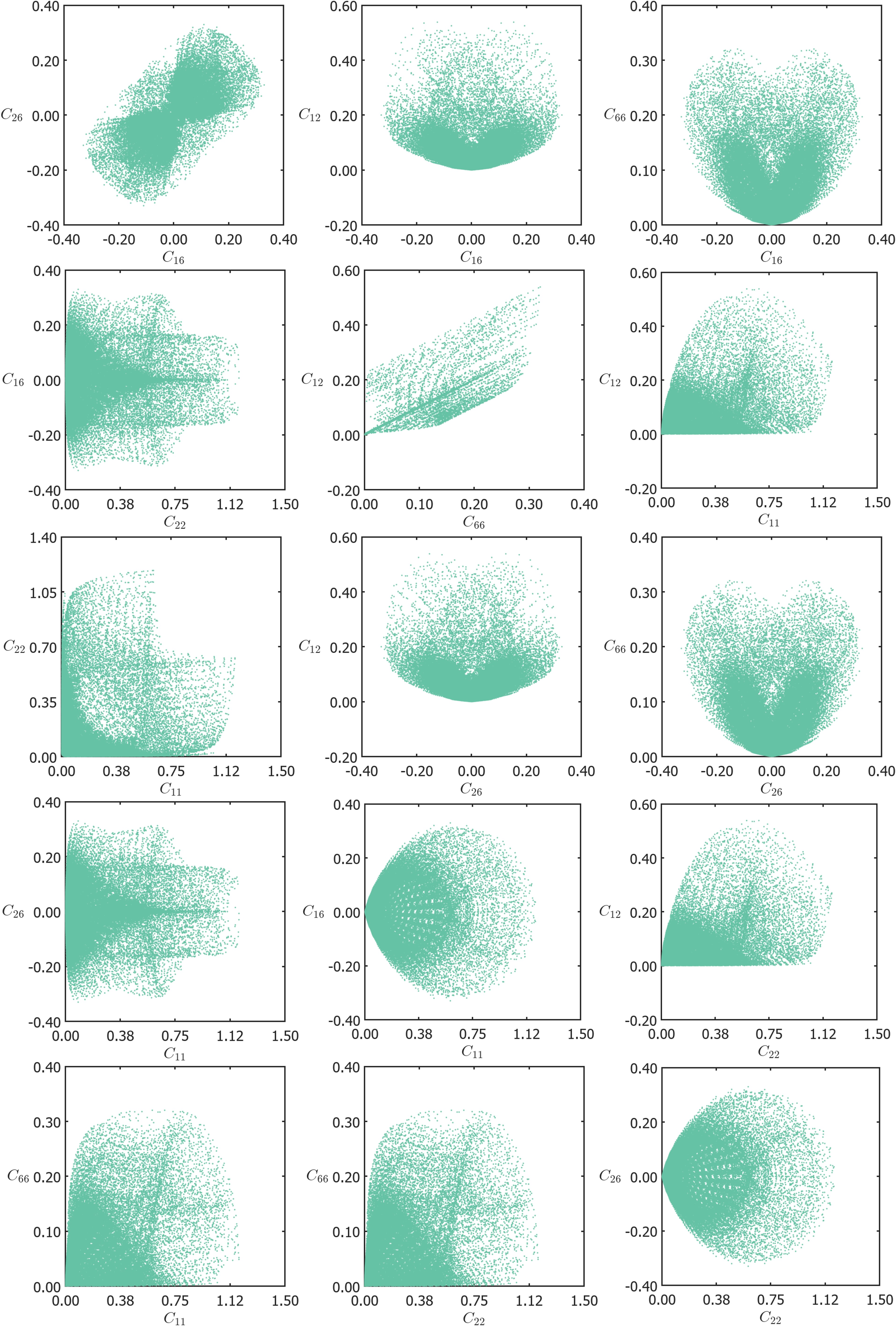}
    \caption{Pair property plots for rank-2 laminates including coordinate transformations. Note that the reach in the properties is significantly increased from rank-1 laminates to rank-2 laminates. This can be clearly observed in the plots of $C_{16}$ vs. $C_{26}$, $C_{11}$ vs. $C_{22}$, $C_{66}$ vs. $C_{12}$. }\label{fig: Cij_vs_Ckl_Rk2}
    \end{center}
\end{figure}

\begin{figure}[!htb]
    \begin{center}
    \centering 
    \includegraphics[width = 0.8\textwidth]{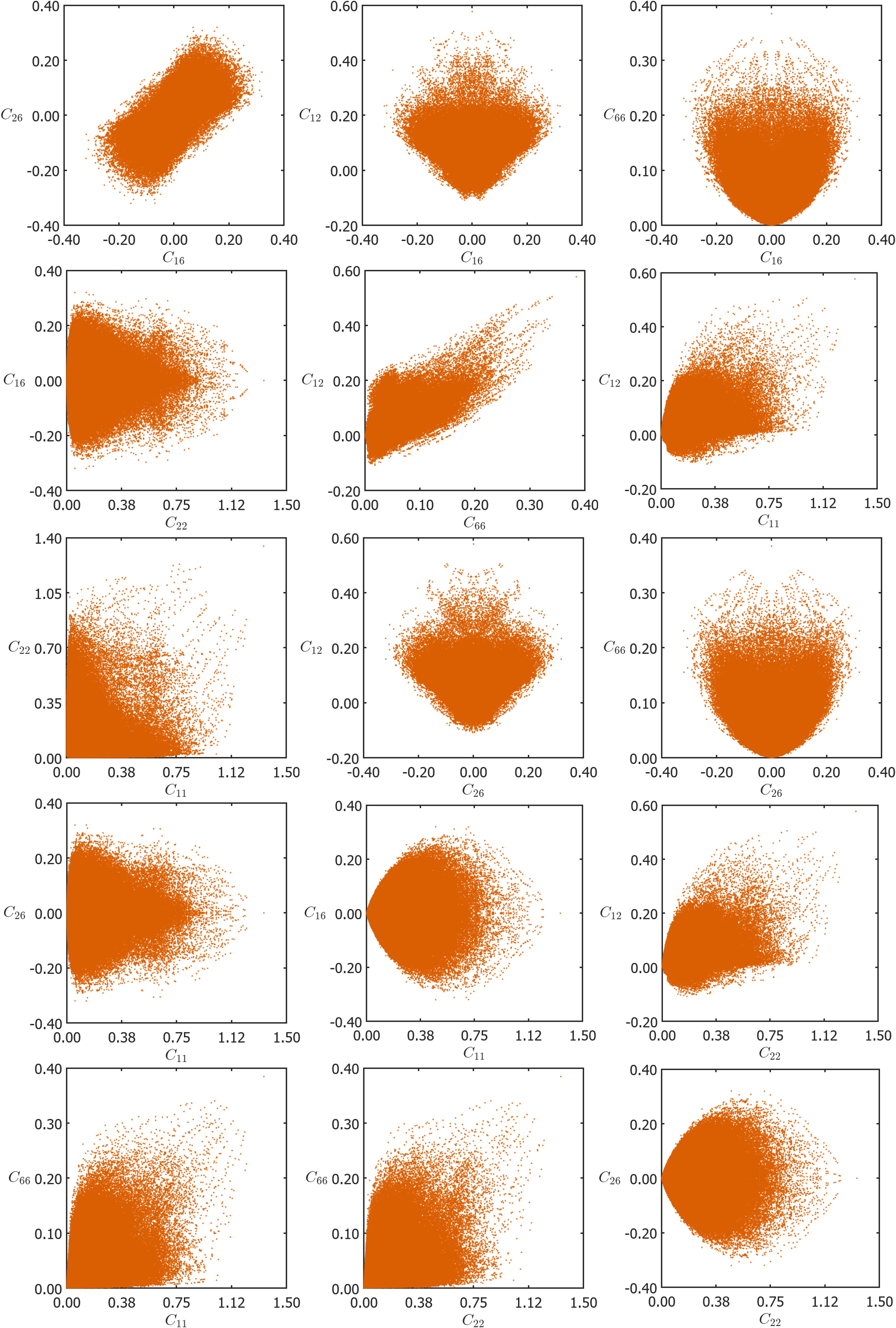}
    \caption{Pair property plots for rank-3 laminates including all possible coordinate transformations generated by using rank-2 laminates as the constituent materials of the two-phase rank-1 lamiantes.}\label{fig: Cij_vs_Ckl_Rk3}
    \end{center}
\end{figure}

\begin{figure}[!htb]
    \begin{center}
    \centering 
    \includegraphics[width = 0.9\textwidth]{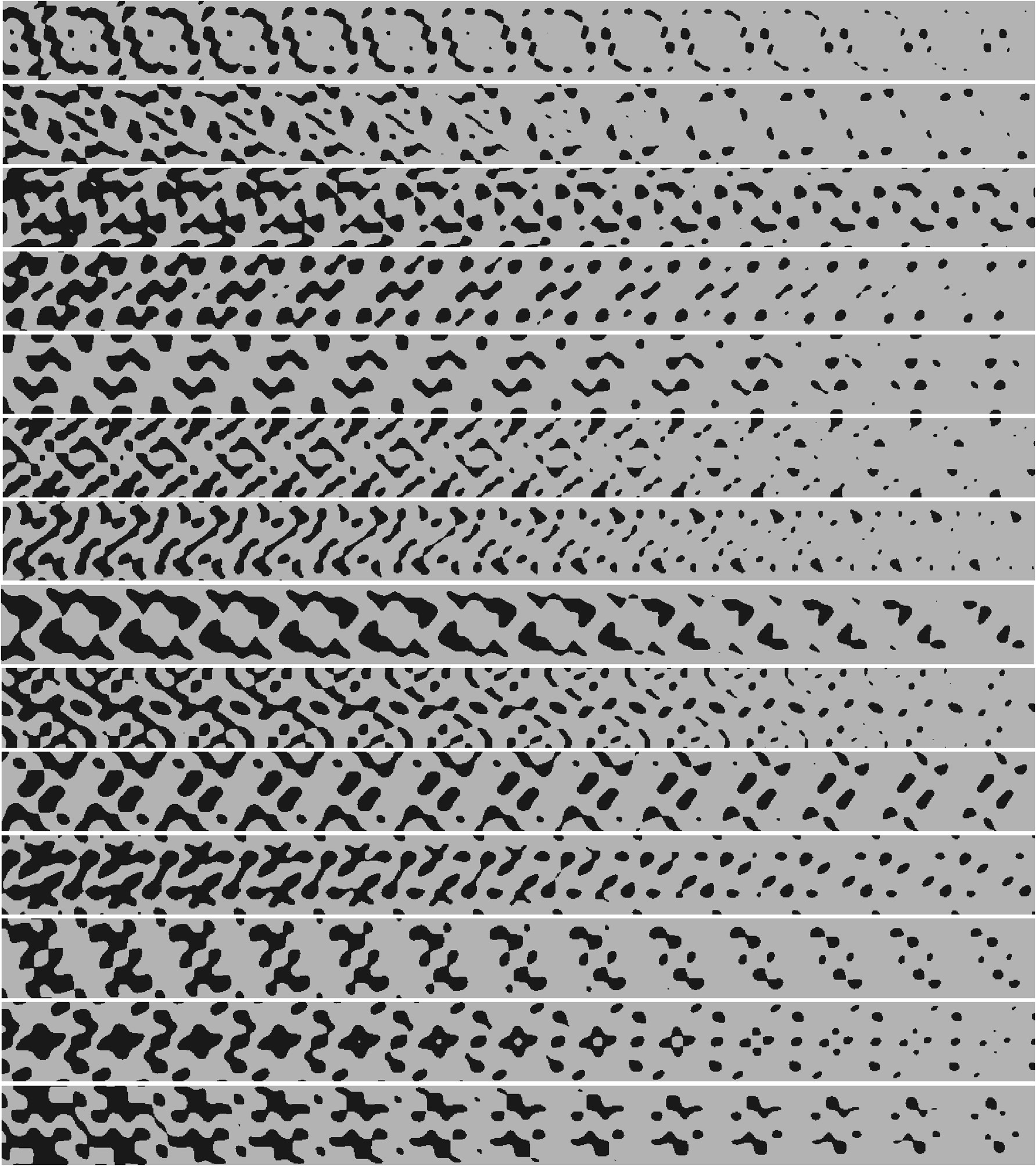}
    \caption{Examples of functionally graded structures with increase in fill fraction from left to right. Each of these gradients are generated from a fixed function while increasing the threshold value of the function.}\label{fig: Gradients 1D Fixed Function}
    \end{center}
\end{figure}

\begin{figure}[!htb]
    \begin{center}
    \centering 
    \includegraphics[width = 0.9\textwidth]{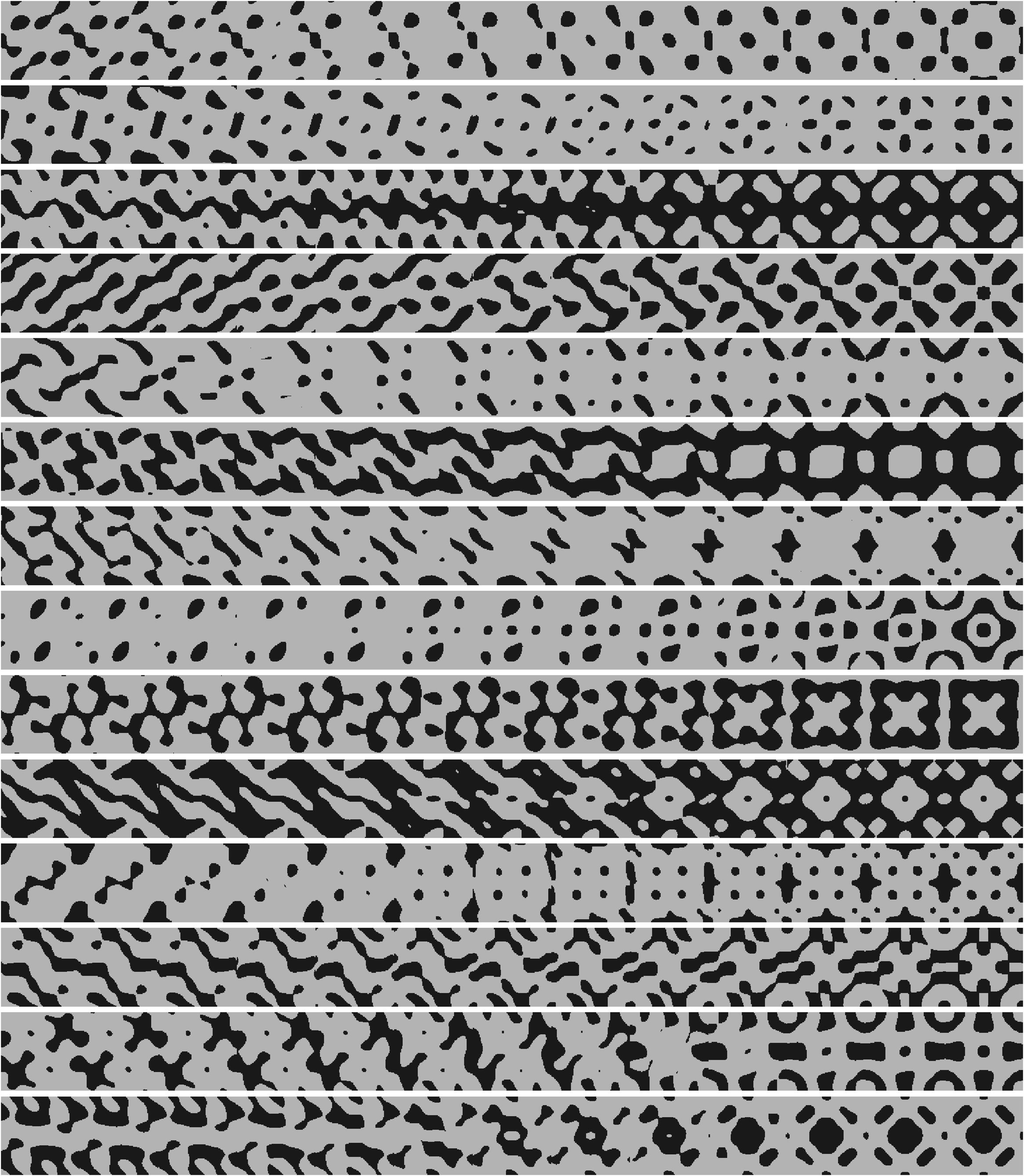}
    \caption{Examples of functionally graded structures with interpolation between unit cells that transition from p2 symmetry to p4 symmetry from left to right.}\label{fig: Gradients 1D Asymmetry to Symmetry}
    \end{center}
\end{figure}

\begin{figure}[!htb]
    \begin{center}
    \centering 
    \includegraphics[width = 0.9\textwidth]{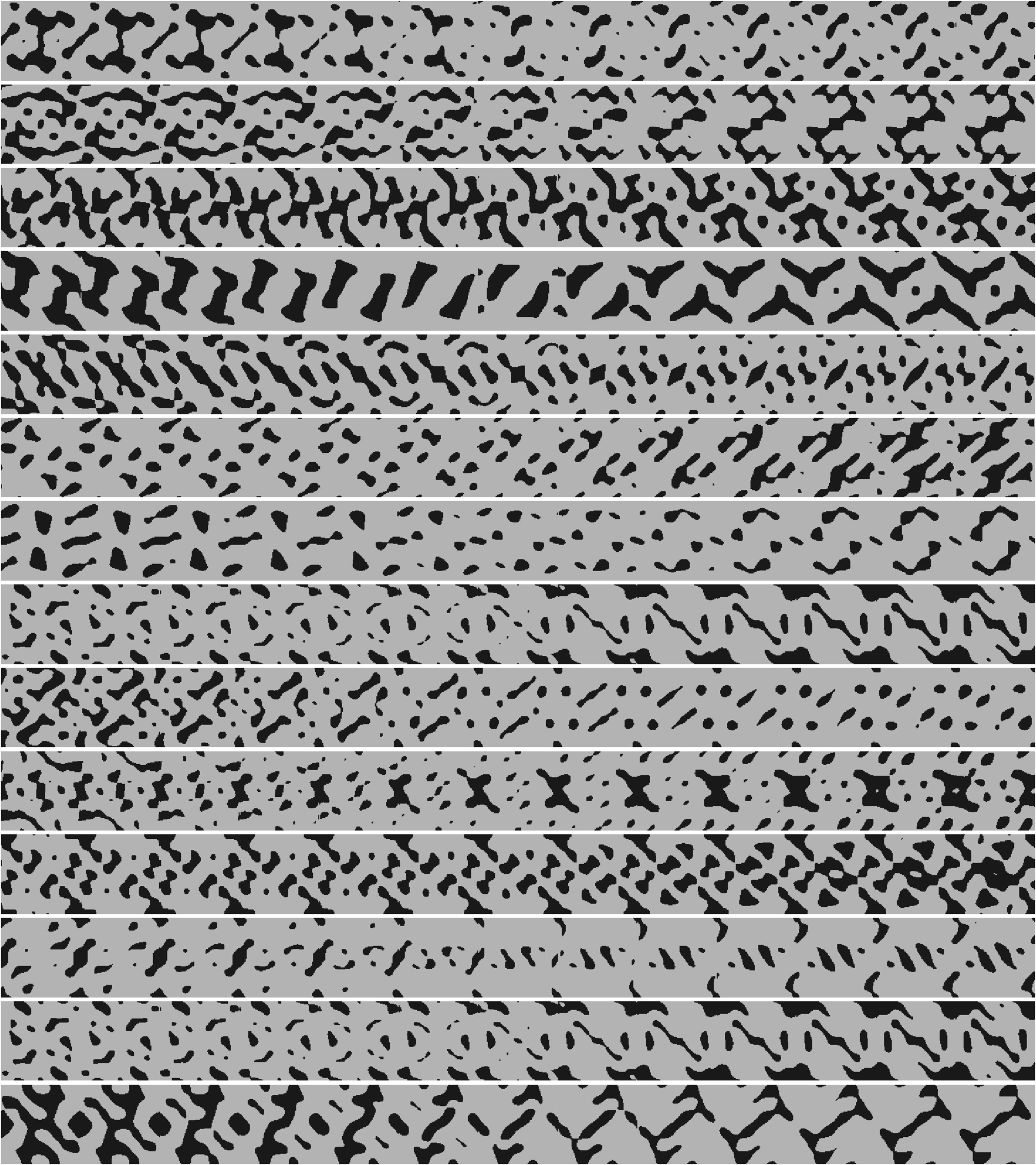}
    \caption{Examples of functionally graded structures with interpolation between two unit cells of p2 symmetry.}\label{fig: Gradients 1D Asymmetry to Asymmetry}
    \end{center}
\end{figure}

\begin{figure}[!htb]
    \begin{center}
    \centering 
    \includegraphics[width = 0.9\textwidth]{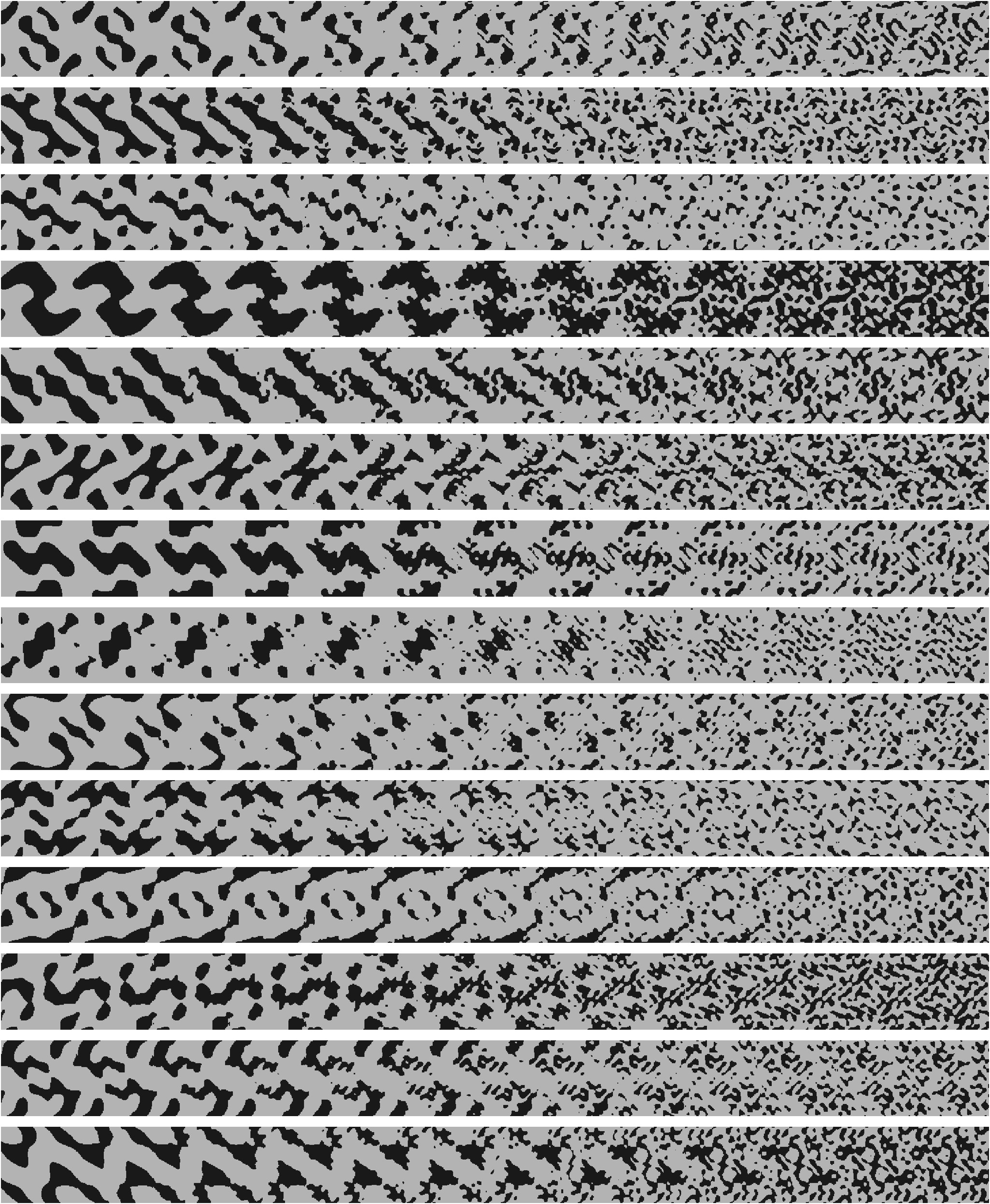}
    \caption{Examples of functionally graded structures with interpolation between unit cells whose number of spatial frequencies increase from left to right.}\label{fig: Gradients Low to high spatial}
    \end{center}
\end{figure}

\begin{figure}[!htb]
    \begin{center}
    \centering 
    \includegraphics[width = 0.8\textwidth]{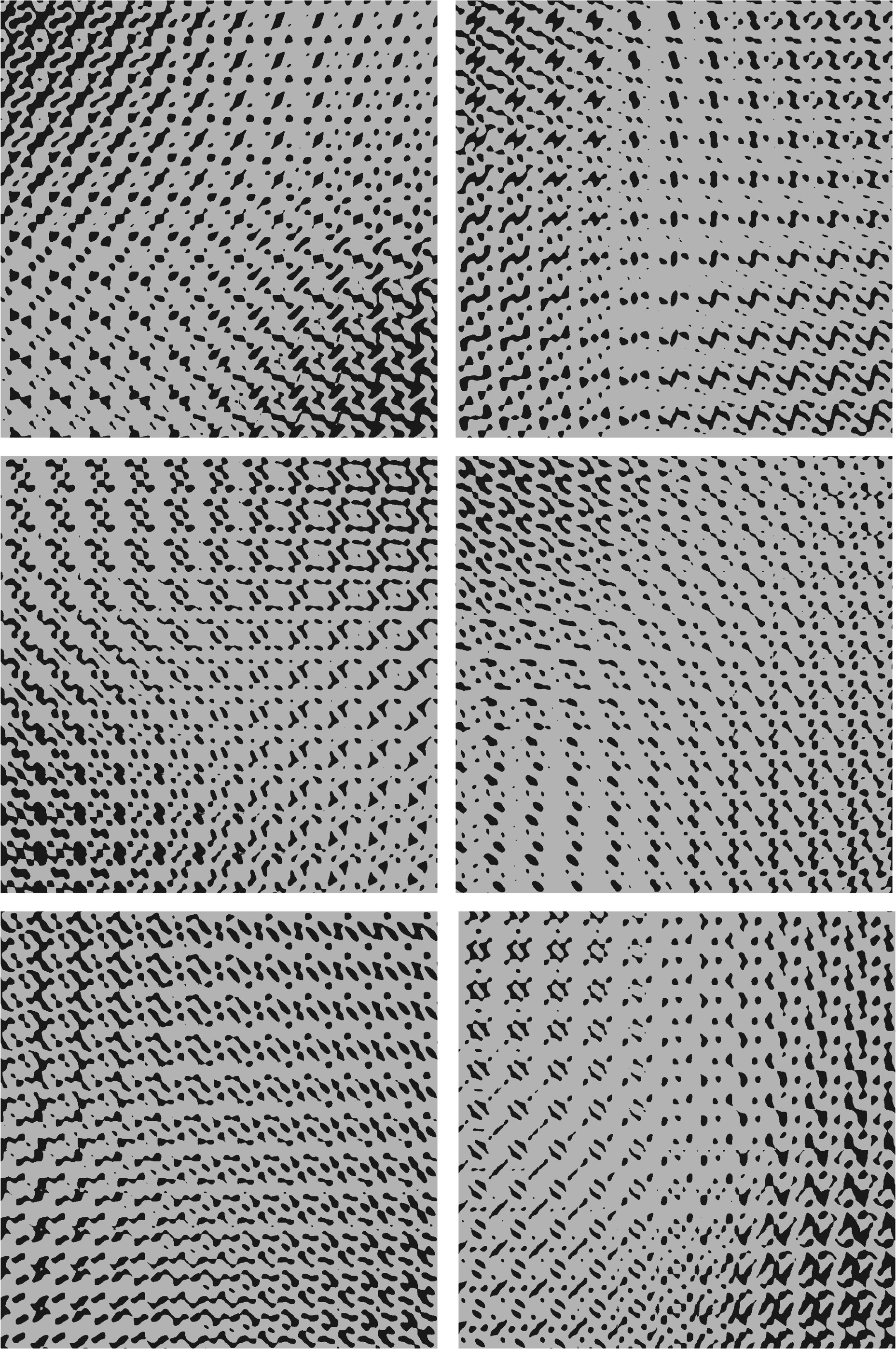}
    \caption{Examples of functionally graded structures with bilinear interpolation between four unit cells with distinct patterns.}\label{fig: Gradients 2D Linear}
    \end{center}
\end{figure}

\begin{figure}[!htb]
    \begin{center}
    \centering 
    \includegraphics[width = 0.9\textwidth]{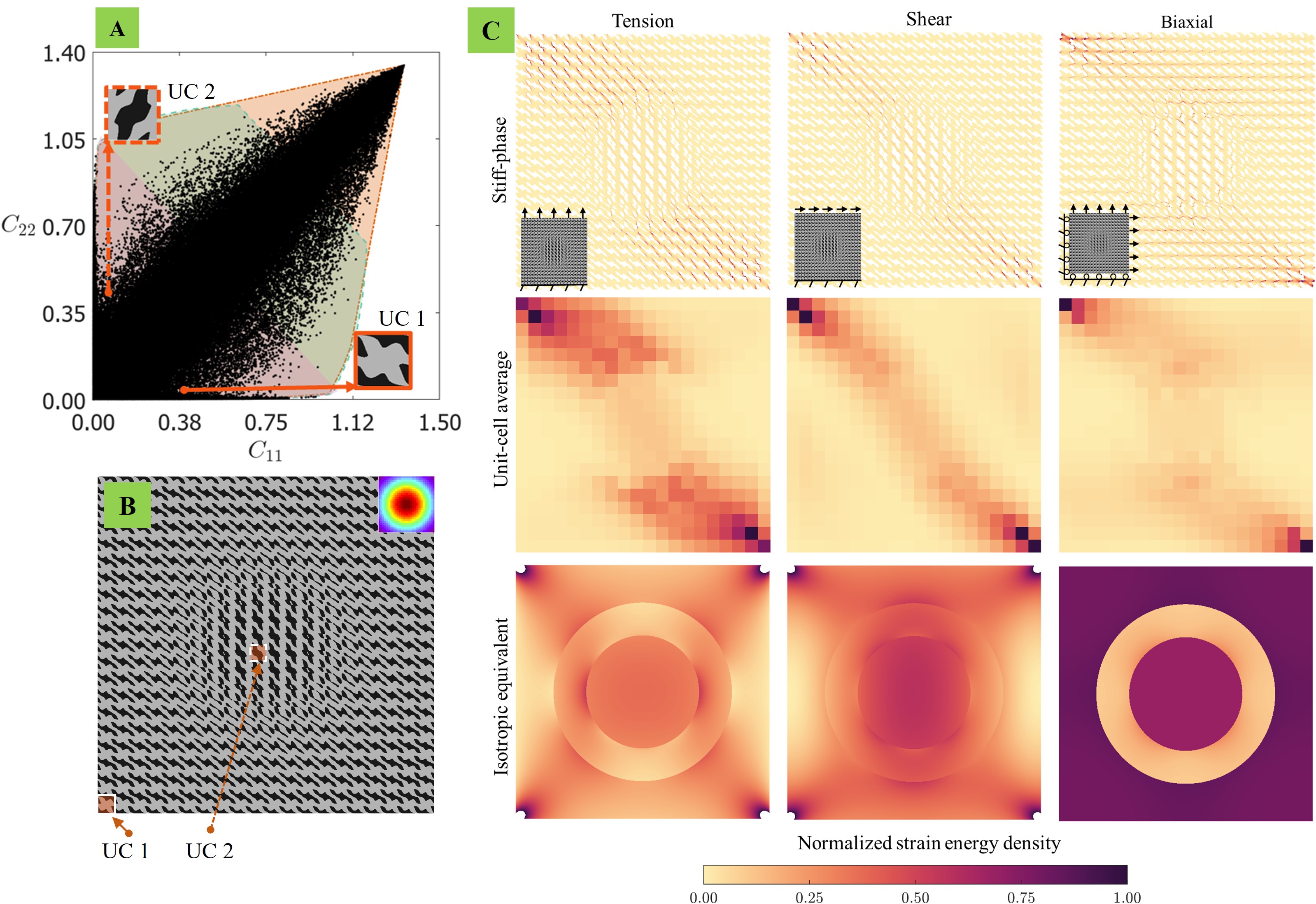}
    \caption{Another demonstration of selective energy localization in radially graded structure akin to \cref{fig: Selective Energy Localization}. (A) Unit cell selection based on the extremity in the property space plot of $C_{11}$ vs. $C_{22}$. The elasticity tensor of the unit cell (named UC 1) at the boundary is [0.378, 0.066, 0.068, -0.083, -0.041, 0.067]$^T$. The elasticity tensor of the unit cell (named UC 2) at the interior is [0.116, 0.086, 0.479, -0.077, -0.078, 0.097]$^T$. The fill fractions of the stiff phase in the unit cells are [59.9\%, 59.3\%] respectively. (B)  Radially graded design with 20 $\times$ 20 tessellation from the chosen unit cells UC 1, UC 2. (C) Distribution of (normalized) elastic energy stored in the circularly interpolated structure for tensile, shear, and biaxial loading displaying selective energy localization arising from anisotropy of the unit cells. Although the fill fractions of the two unit cells are almost the same, the radial interpolation resulted in unit cells with higher fill fractions in the interface part of the graded region. Therefore, an approximate annular region filled with a stiffer isotropic medium is used when calculating the isotropic equivalence.}\label{fig: EnergyLocalizationSample2}
    \end{center}
\end{figure}

\begin{figure}[!htb]
    \begin{center}
    \centering 
    \includegraphics[width = \textwidth]{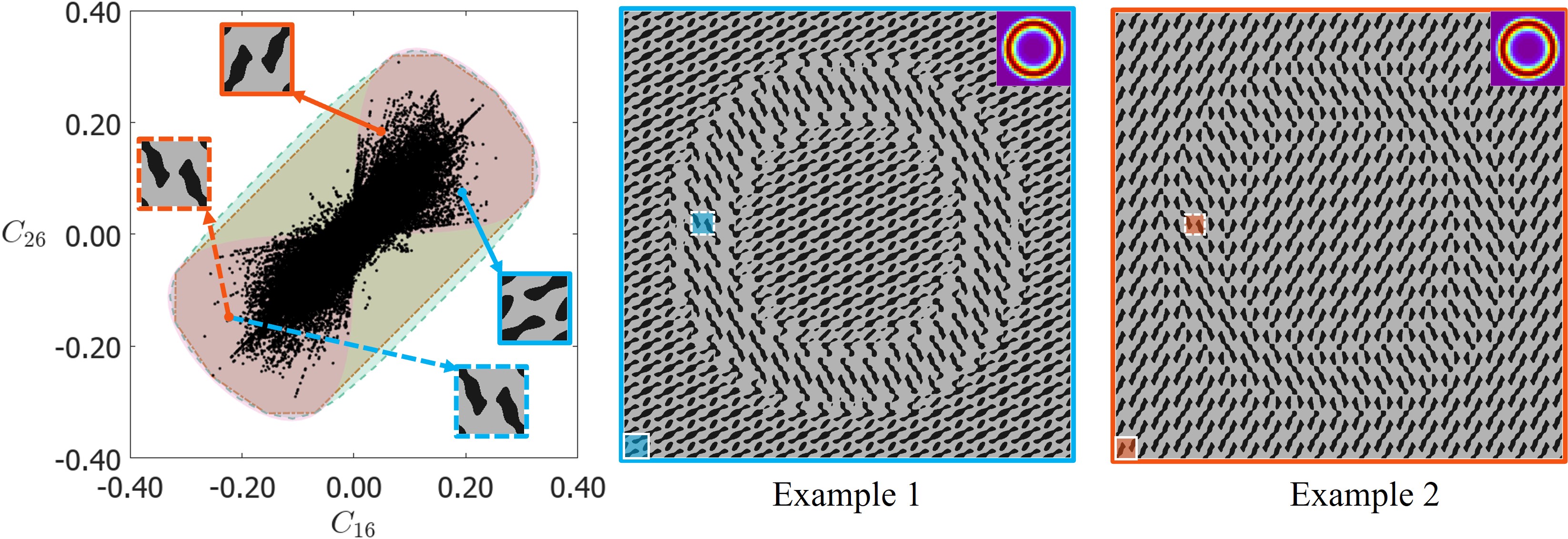}
    \caption{The selection of unit cells used for annular interpolation is based on the extremity of the property plot of $C_{16}$ vs. $C_{26}$. The annular interpolations in examples 1 and 2 illustrate non-affine rotation-like deformation under tension in \cref{fig: Selective Strain Localization 1,fig: Selective Strain Localization 2}. Note that in example 2, the unit cell in the annular region is obtained by a 90$^\circ$ rotation of the unit cell at the boundary. Additionally, it is important to mention that these designs can be fabricated using only the stiff phase.}\label{fig: Strain_Localization_Geometries}
    \end{center}
\end{figure}

\begin{figure}[!htb]
    \begin{center}
    \centering 
    \includegraphics[width = 0.8\textwidth]{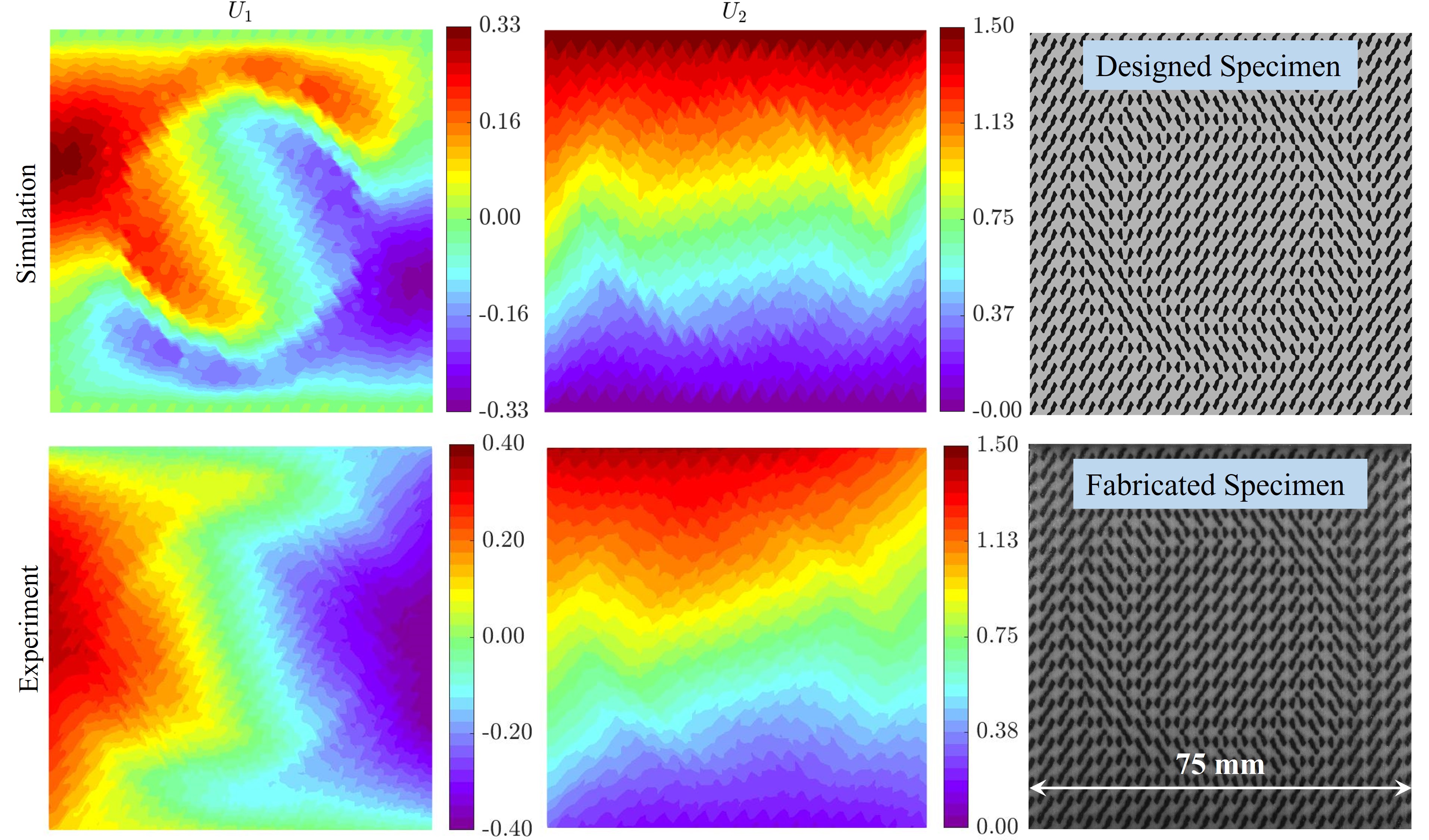}
    \caption{Rotation-like deformation under tensile loading in the gradient structure named example 2 as shown in \cref{fig: Strain_Localization_Geometries} made with annular interpolation. The geometric incompatibility between two unit cells with opposing shear-normal coupling behavior leads to non-affine deformation. The top row shows finite element simulation results while the bottom row compares the displacement contours measured using the digital image correlation (DIC) on an additively manufactured specimen. }\label{fig: Selective Strain Localization 2} 
    \end{center}
\end{figure}

 \newpage
 \begin{figure}[!htb]
	\begin{center}
		\centering 
		\includegraphics[width = 0.8\textwidth]{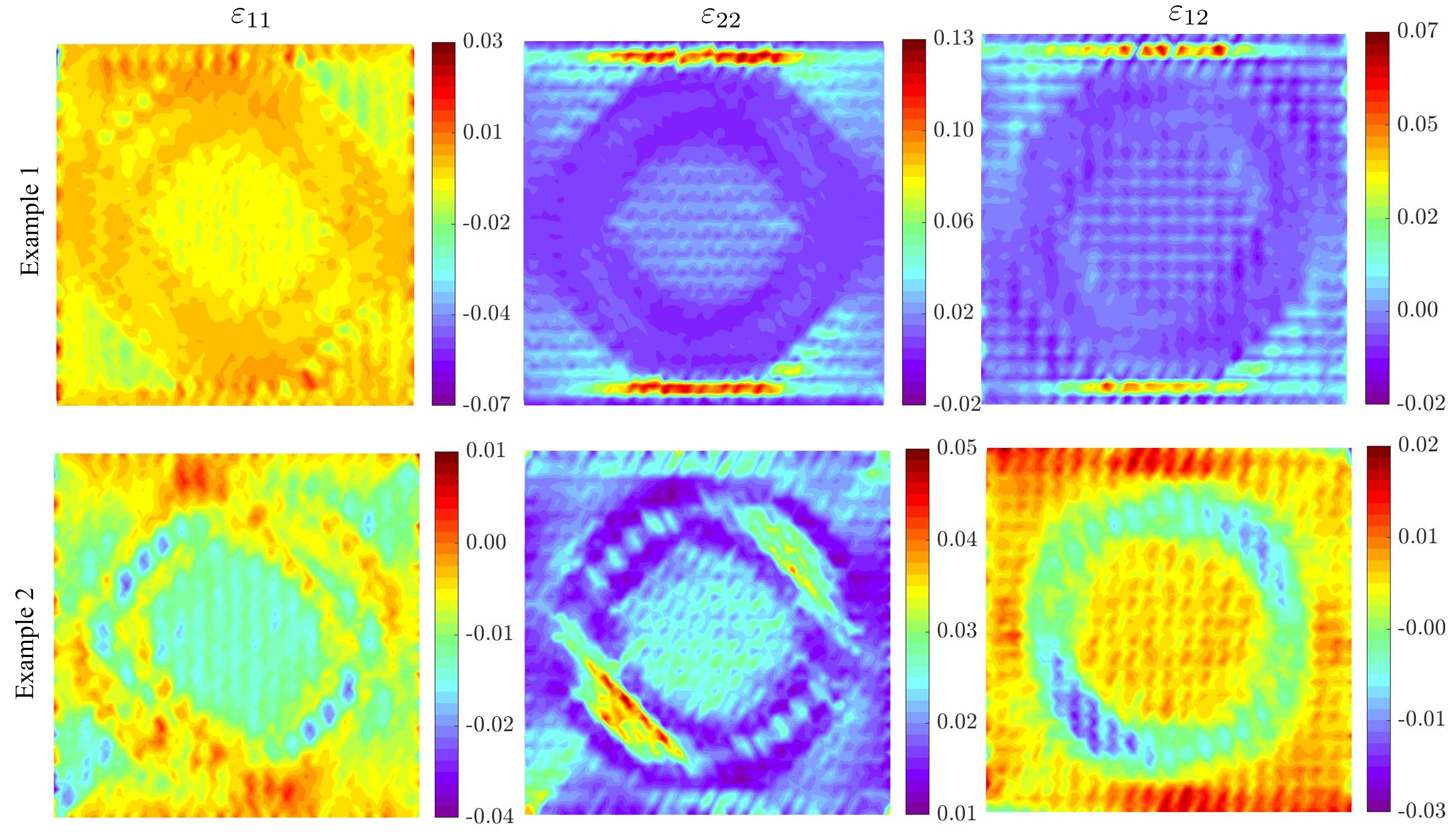}
		\caption{Experimentally measured strains obtained using DIC for example 1 shown in \cref{fig: Selective Strain Localization 1} and for example 2 shown in \cref{fig: Selective Strain Localization 2} that discuss non-affine deformations. In both examples, the strain contours indicate strain localization in the annular region which is different from the strain observed in the rest of the structure. Stresses are not plotted as they are difficult to obtain experimentally.}	\label{fig: Selective Strain Localization Strains}
	\end{center}
\end{figure} 

\begin{figure}[!htb]
	\begin{center}
		\centering 
		\includegraphics[width = 0.9\textwidth]{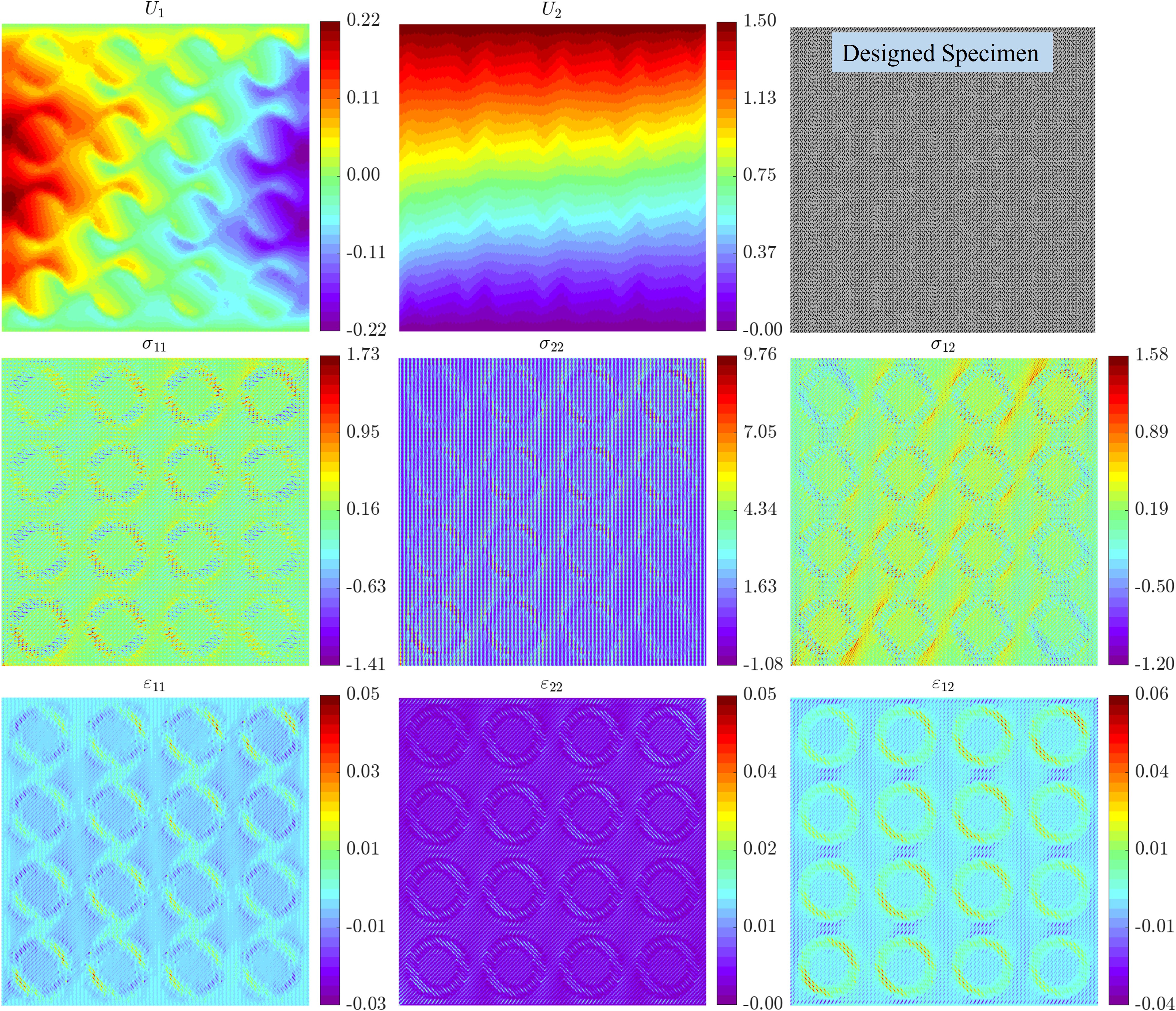}
		\caption{Mechanical behavior in a $ 4\times 4$ {supercell} tessellation design subjected to tensile loading. The {supercell} consists of unit cells with opposing shear-normal coupling arranged in an annular interpolation scheme, which is the entire specimen considered in \cref{fig: Selective Strain Localization 2}. The displacement contour $U_1$ displays multiple regions of rotation-like deformation arising from incompatibilities in the deformation modes of the unit cells. $\sigma_{11},\sigma_{22}$ contours {(in units of MPa)} display how these incompatibilities in {$C_{16}, C_{26}$ lead to alternative regions of compressive and tensile stresses in the tessellated supercell} undergoing tensile loading. $\sigma_{12}$ contour shows the rotation-induced shear stress localization. All the strain contours further corroborate the localization of the strains around the annular interface.}	\label{fig: Selective Strain Localization Sample2 4by4}
	\end{center}
\end{figure}

%%%%%%%%%%%%%%%%%%%%%%%%%%%%%%%%%%%%%%%%%
%%%%%%%%%%%%%% THE END %%%%%%%%%%%%%%%%%%
%%%%%%%%%%%%%%%%%%%%%%%%%%%%%%%%%%%%%%%%%
\end{document}